\magnification1200


\vskip 2cm
\centerline
{\bf  E11, Romans theory and higher level duality relations}
\vskip 1cm
\centerline{ Alexander G. Tumanov and Peter West}
\centerline{Department of Mathematics}
\centerline{King's College, London WC2R 2LS, UK}
\vskip 2cm
\leftline{\sl Abstract}
From the underlying non-linear realisation we  compute the complete $E_{11}$ invariant equations of motion in eleven dimensions, at the linearised level,  up to and including  level four in the  fields. Thus we include the   metric, the three and six forms, the dual graviton and three fields at level four. The fields are linked by a set of duality equations, which are first order in derivatives and  transform into each other under the $E_{11}$ symmetries. From these duality relations we deduce second order equations of motion,  including  those for the usual supergravity fields.   As a result  the on-shell degrees of freedom are those of the eleven dimensional supergravity. We also show that the level four fields provide an eleven dimensional origin of Romans theory  and lead to a novel duality relation.

\vskip2cm
\noindent

\vskip .5cm

\vfill
\eject

\medskip 
{\bf 1. Introduction}
\medskip
The non-linear realisation of the semi-direct product of  $E_{11}$ and its vector representation leads to an $E_{11}$ invariant field theory that has an infinite number of fields which live in a space time that has an infinite number of coordinates [1,2]. These fields and coordinates can each be classified by a level. By taking different decomposition of $E_{11}$ into the subalgebras $GL(D)\otimes E_{11-D}$ one finds theories in $D$ dimensions [3,4,5,6]. The fields at low levels are just those of the maximal supergravity theories in the different dimensions and the lowest level coordinates are just the coordinates of our usual space times.  The non-linear realisation also determines  the equations of motion that the fields obey. The final result is a set of field equations which rotate into each other under the $E_{11}$ symmetry. 
\par
It was recently shown that the non-linear equations of motion of the eleven dimensional theory are uniquely determined  at low levels and they are precisely those of eleven dimensional supergravity [7,8]. The analogous calculation has been carried out in five dimensions [7] and,  taking into account the results in eleven dimensions, one finds the same conclusion [8]. It is inevitable that similar results apply in all the other dimensions. This essentially confirms the $E_{11}$ conjecture, namely that the low energy effective action of strings and branes has an $E_{11}$ symmetry. 
\par
In eleven dimensions the fields at level zero one and two are the metric ($h_a{}^b$), the three form ($A_{a_1a_2a_3}$) and the six form ($A_{a_1\ldots a_6}$) respectively, at level  three we find the dual graviton ($h_{a_1\ldots a_8, b}$) and at level four we have the fields [9]
$$
A_{a_1\ldots a_9, b_1b_2b_3}, \ \ A_ {a_1\ldots a_{10}, b_1b_2} , \ \ 
A_{a_1\ldots a_{11}, b}
\eqno(1.1)$$
In this equation the indices in a given block are antisymmetric except for the second block for  the field $A_ {a_1\ldots a_{10}, b_1b_2}$ which is symmetric, that is, $A_ {a_1\ldots a_{10}, b_1b_2}= A_ {a_1\ldots a_{10}, (b_1b_2)}$. The fields also obey the usual SL(11) irreducibility conditions, that is, $A_{[ a_1\ldots a_9, b_1 ] b_2b_3}=0$ and  $A_ {[ a_1\ldots a_{10}, b_1 ] b_2}=0 $. 
It is very well known that the six form $A_{a_1\ldots a_6}$ is an alternative way of describing the degrees of freedom of the three form $A_{a_1a_2a_3}$ and one can write down a duality equation that relates the two field strengths. In a similar way it had been conjectured that the first field in equation (1.1), that is  $A_{a_1\ldots a_9, b_1b_2b_3} $,  is also  an alternative way of describing the degrees of freedom usually encoded in the three form field [10]. Indeed at even higher level of $E_{11}$ one finds  fields which possess  three antisymmetrised indices  as well as  arbitrary numbers of blocks of nine antisymmetrised indices, that is, fields of the form  $A_{a_1\ldots a_9,b_1\ldots b_9, \ldots, c_1c_2c_3}$ as well as  similar fields but with the three antisymmetrised indices replaced by six   indices, that is $A_{a_1\ldots a_9,b_1\ldots b_9, \ldots , c_1\ldots c_6} $ [10]. Motivated by the presence of these fields duality equations that are first order in derivatives and which relate the field $A_{a_1a_2a_3}$ to the field $A_{a_1\ldots a_9, b_1b_2b_3}$, as well as similar equations for the higher level  fields,  were derived in reference [11] using just the knowledge of the irreducible representations of the Poincare group. 
The second field in equation (1.1) has been conjectured to lead, when dimensionally reduced, to the Romans theory [12] in the context of the ten dimensional IIA supergravity [13]. In particular the field $A_ {a_1\ldots a_9 11, (11 11)}$ is the desired nine form that leads to a cosmological constant. It will turn out that the third field in equation (1.1) does not enter the dynamics when it is restricted to contain only the usual spacetime derivatives. 
\par
In this paper we will derive the dynamical equations in eleven dimensions  of the $E_{11}$ non-linear realisation up to and including the above level four fields. We will do this at the linearised level. The results can be summarised in the table one below. 

\bigskip
{\bf Table 1. The $E_{11}$ variations of the duality relations and equations of motion}
\medskip
$${\normalbaselineskip=20pt \matrix{ 
E^{(1)}{}_{a_1a_2a_3a_4}=0 & \Leftrightarrow & E^{(1)}{}_a{}^{b_1b_2} \dot =0 & \Leftrightarrow & E^{(1)}{}_{a_1\ldots a_{10}, b_1b_2b_3}\dot =0 , \ E^{(1)}{}_{a_1\ldots a_{11}, b_1b_2}\dot =0\cr
\Downarrow & ÊÊ& \Downarrow Ê& ÊÊ& \Downarrow \cr 
E^{(2)}{}_{a_1a_2a_3}=0 Ê& ÊÊ& E^{(2)}{}_a{}^b=0 Ê& Ê& E^{(2)}{}_{a_1\ldots a_9 , b_1b_2b_3}=0 Ê\cr
& Ê\Leftrightarrow & Ê& \Leftrightarrow & \cr
ÊE^{(2)}{}_{a_1\ldots a_6}=0 Ê& ÊÊ& ÊE^{(2)}{}_{a_1\ldots a_8,b} =0 & & ÊE^{(2)}{}_{a_1\ldots a_{11}, b_1b_2, c}\dot =0 \cr
& & & &\Downarrow \cr
& & & & E^{(3)}{}_{a_1\ldots a_{11}, b_1b_2, c_1 c_2}=0  \cr\cr}}
$$
\bigskip
The different sections of the paper consist of carrying out the steps shown in the above table. The top row consists of duality relations which contain only one space time derivative. The numerical superscript on the symbol $E$ indicates the number of spacetime derivatives. 
The first equation in the top  row is the familiar, and previously derived for $E_{11}$ [14,8], duality relation $E^{(1)}{}_{a_1\ldots a_4}=0$ between the three form, $A_{a_1a_2a_3}$ and six form $A_{a_1\ldots a_6}$ fields. Its $E_{11}$ variation represented by  $\Leftrightarrow$ leads to the duality relation $E^{(1)}{}_a{}^{b_1b_2} \dot =0$ between the usual gravity field $h_{ab}$ and the dual gravity field $h_{a_1\ldots a_8 , b}$. The $E_{11}$ variation of this last duality relation leads to two relations 
$E^{(1)}{}_{a_1\ldots a_{10}, b_1b_2b_3}\dot =0$ and $ E^{(1)}{}_{a_1\ldots a_{11}, b_1b_2}\dot =0$. The first of these gives a duality relation between the field $A_{a_1\ldots a_9, b_1b_2b_3}$ and the three from $A_{a_1a_2a_3}$ while the second, unlike the others,  just involves the one field $A_{a_1\dots a_{10}, b_1b_2}$. 
\par
From each of the  equations  just mentioned, which are first order in derivatives, one can deduce equations of motion that are second order in derivatives which are indicated by the $\Downarrow$ arrow and occur in the second  row of the table below. From the $E^{(1)}{}_{a_1\ldots a_4}=0$ duality relation we deduce the usual field equations for the three forms and six form, that is, $E^{(2)}{}^{a_1a_2a_3}=0$ and  $ E^{(2)}{}^{a_1\ldots a_6}=0 $. From the gravity- dual gravity duality relation $E^{(1)}{}_a{}^{b_1b_2} \dot =0$ we find the linearised Einstein equation, that is, $E^{(2)}{}_a{}^b=0$ and the equation of motion for the dual graviton, that is,  $E^{(2)}{}_{a_1\ldots a_8,b}=0$. While from the duality relation $E^{(1)}{}_{a_1\ldots a_{10}, b_1b_2b_3}\dot =0$ one can deduce the  relation $E^{(2)}{}_{a_1\ldots a_{10} , b_1b_2b_3b_4}=0$ which is second order in derivatives and relates the field strength for the field $A_{a_1\ldots a_9, b_1b_2b_3}$ to the derivatives of the field strength for the three form $A_{a_1a_2a_3}$. Taking the trace we find the object $E^{(2)}{}_{a_1\ldots a_9 , b_1b_2b_3}\equiv  E^{(2)}{}_{c a_1\ldots a_{9} , c b_1b_2b_3}=0$ which appears in the table as it is this object that one finds in the $E_{11}$ variation of $E^{(2)}{}_{a_1\ldots a_8,b}$. In fact upon taking three more traces one eliminates the three  form $A_{a_1a_2a_3}$ to find the correct equation of motion for the  field $A_{a_1\ldots a_9, b_1b_2b_3}$ although this step is not shown in the table. 
\par
The use of the symbol $\dot =$ means that the equation holds modulo certain transformations which are specified later in the paper.  However, we note that if one takes sufficient derivatives,  in an appropriate way, of the  equations that hold modulo certain transformations one finds equations that hold in the usual sense. Thus $E_{11}$ gives rise to a web of equations which include the usual equations of motion we are familiar with. 
\par
We also carry out the $E_{11}$ variations of the second order equations of motion as indicated by the $\Leftrightarrow$ in the second and third  rows of the table. As displayed one finds that the equations of motion vary into each other.
\par
By taking derivatives of the equation $E^{(1)}{}_{a_1\ldots a_{11}, b_1b_2}$
for the field $A_{a_1\ldots a_{10}, b_1b_2}$, which is first order in derivatives one finds  the equation $E^{(2)}{}_{a_1\ldots a_{11}, b_1b_2, c}\dot =0$ which also only holds modulo certain transformations. As indicated  in the table one can take one more derivative to find the equation $E^{(3)}{}_{a_1\ldots a_{11}, b_1b_2, c_1 c_2}=0 $ in the fourth row which holds exactly. The $E_{11}$ variation of this last equation results in the previous equations, although this fact is  not shown in the table, 
\par

\medskip 
{\bf 2. The construction of the non-linear realisation }
\medskip
The construction of the non-linear realisation starts from the group element $g\in E_{11}\otimes_S l_1$ which is subject to the transformations $g\to g_0 gh$ where $g_0 \in E_{11}\otimes_S l_1$ is a rigid transformation and $h\in I_c(E_{11})$ is a local transformation. This leads to a field theory equipped with equations of motion that are invariant under the $E_{11}$ symmetries of the non-linear realisation. Discussions of how this works can be found in many $E_{11}$ papers, see for example [1,14] and the review [15]. 
\par
We can write the group element $g$ in the form  $g= g_l g_E$ and parmeterise it as follows  
$$
g_E=  \ldots e^{R^{c_1\ldots c_{11},b} A_{c_1\ldots c_{11},b} }e^{ R^{c_1\ldots c_{10},b_1 b_2}A_{c_1\ldots c_{10},b_1 b_2}}e^{ R^{c_1\ldots c_9,b_1 b_2b_3} 
A_{c_1\ldots c_9,b_1 b_2b_3}} 
$$
$$
\times e^{ h_{a_1\ldots a_{8},b}
R^{a_1\ldots a_{8},b}} e^{ A_{a_1\ldots
a_6} R^{a_1\ldots a_6}}e^{ A_{a_1 a_2 a_3} R^{a_1 a_2
a_3}} e^{h_a{}^b K^a{}_b}\equiv e^{A_{\underline \alpha} R^{\underline \alpha}}
\eqno(2.1)$$ 
and 
$$
g_l= e^{x^aP_a} e^{x_{ab}Z^{ab}} e^{x_{a_1\ldots a_5}Z^{a_1\ldots a_5}}\ldots = 
e^{z^A l_A} 
\eqno(2.2)$$
In these equations  $R^{\underline \alpha}$ are the generators of $E_{11}$ and $l_A$ the generators in the $l_1$ representation. The group element $g_E$ contains the fields $A_{\underline \alpha}$ and it extends the previous formulations [14] to include the level four generators and fields. 
The blocks of indices on  the generators are totally antisymmetrised  except for the generator $R^{c_1\ldots c_9,b_1b_2}$ which obeys $R^{c_1\ldots c_9,b_1b_2}=R^{c_1\ldots c_9,(b_1b_2)}$. They all obey the constraints corresponding to   irreducible representation of SL(11), in particular
$$
R^{[c_1\ldots c_8,b]}=0=R^{[c_1\ldots c_9,b_1 ]b_2b_3}=R^{[c_1\ldots c_{10},b_1 ] b_2}
\eqno(2.3)$$
The group element $g_l$ is parameterised by the   $z^A$ upon which the fields depend and so are the coordinates of the spacetime.  
\par
To construct the dynamics we will use the Cartan forms which are defined by 
$$
{\cal V}= g^{-1} dg = {\cal V}_E +{\cal V}_l, 
\eqno(2.4)$$
where 
$$
{\cal V}_E=g_E^{-1}dg_E\equiv dz^\Pi G_{\Pi, \underline \alpha} R^{\underline \alpha}, \ {\rm and }\ 
{\cal V}_l= g_E^{-1}(g_l^{-1}dg_l) g_E= g_E^{-1} dz\cdot l g_E\equiv 
dz^\Pi E_\Pi{}^A l_A  
 \eqno(2.5)$$
\par
The Cartan form  ${\cal V}_E$ belongs to the $E_{11}$ algebra and they can be expressed as 
$$
{\cal V}_E= 
G _{a}{}^b K^a{}_b+  G_{c_1\ldots c_3}
R^{c_1\ldots c_3} +G_{c_1\ldots c_6} R^{c_1\ldots c_6}+
G_{c_1\ldots c_8,b} R^{c_1\ldots c_8,b}
$$
$$
+G_{c_1\ldots c_9,b_1b_2b_3}R^{c_1\ldots c_9,b_1b_2b_3}+ G_{c_1\ldots c_{10},b_1b_2}R^{c_1\ldots c_{10},b_1b_2} +G_{c_1\ldots c_{11},b}R^{c_{11}\ldots c_{11},b}+\ldots 
\eqno(2.6)$$

\par
The Cartan form ${\cal V}_l$ is in the space of generators of the $l_1$ representation and one can recognise $E_\Pi{}^A = (e^{A_{\underline \alpha}D^{\underline \alpha}})_\Pi{}^A$ as the vielbein on the generalised space time.

\par
In this paper we will work at  the linearised level, that is to first order in the fields. In this approximation   the Cartan forms are given by [14]
$$
G_{a,b}{}^{c}= \partial_{a}h_{b}{}^{c}, \ 
G_{a_1,a_2a_3a_4 }= \partial_{a_1} A_{a_2a_3a_4}, 
G_{a_1,\ldots a_7 }=\partial_{a_1} A_{a_2\ldots a_7} 
,\  G_{a_1,a_2\ldots a_9, c }= \partial_{a_1}h_{a_2\ldots a_9, c }, 
$$
$$
G_{a_1,\,a_2...a_{10},b_1b_2b_3}= \partial_{a_1}A_{a_2...a_{10},b_1b_2b_3} ,\ 
G_{a_1,\,a_2...a_{11},b_1b_2}= \partial_{a_1}A_{a_2...a_{11},b_1b_2}
$$
$$
G_{a_1,\,a_2...a_{12},b}= \partial_{a_1}A_{a_2...a_{12},b}
\eqno(2.7)$$
\par
The Cartan forms, when viewed as forms,  are inert under the rigid transformations but transform under the local $h$ transformations.  At level zero the $I_c(E_{11})$ transformations of  $I_c(E_{11})$  are just the local Lorentz SO(11) transformations that act in the usual way.  At the next level the $I_c(E_{11})$  transforms   involves the $E_{11}$ 
 generators at levels $\pm 1$ and are  of the form 
$$
h=1- \Lambda _{a_1 a_2  a_3 }S^{a_1 a_2  a_3  }, \quad {\rm  where   }\quad
S^{a_1a_2a_3}= R^{a_1a_2a_3}- \eta^{a_1b_1} \eta^{a_2b_2}\ \eta^{a_3b_3} R_{b_1b_2b_3}
\eqno(2.8)$$
Under these transformations the Cartan forms of equation (2.6) change as     
$$
\delta\,{\cal V}_E = \left[S^{a_1a_2a_3}\,\Lambda_{a_1a_2a_3},{\cal V}_E\right] - S^{a_1a_2a_3}\,d\Lambda_{a_1a_2a_3}.
\eqno(2.9)$$ 
The explicit forms of these transformations  are given by  [14]
$$
\delta G_{a}{}^{b}=18 \Lambda^{c_1c_2 b }G_{c_1c_2 a}
-2 \delta_a ^{b}  \Lambda^{c_1c_2 c_3}G_{c_1c_2 c_3},\ 
\eqno(2.10)$$
$$
\delta G_{a_1a_2a_3}=-{5!\over 2} G_{b_1b_2b_3 a_1a_2a_3}
\Lambda^{b_1b_2 b_3} -6G_{(c [a_1 |) } \Lambda_{c}{}_{|a_2a_3]}
\eqno(2.11)$$
$$
\delta G_{a_1\ldots a_6}=2 \Lambda_{[ a_1a_2a_3}G_{a_4a_5a_6 ]}
-8.7.2 G_{b_1b_2b_3 [ a_1\ldots a_5,a_6]}\Lambda^{b_1b_2b_3}
+8.7.2 G_{b_1b_2 a_1\ldots a_5a_6, b_3 }\Lambda^{b_1b_2b_3}
$$
$$
= 2 \Lambda_{[ a_1a_2a_3}G_{a_4a_5a_6 ]}
-8.7.6 G_{b_1b_2b_3 [ a_1\ldots a_5,a_6]}\Lambda^{b_1b_2b_3}
\eqno(2.12)$$
$$
\delta\,G_{a_1...a_8,\,b} = -\,2\,G_{[a_1...a_6}\,\Lambda_{a_7a_8]b} - 2\,G_{[a_1...a_5|b|}\,\Lambda_{a_6a_7a_8]}
$$
$$
-\,440\,\left(G_{a_1...a_8e_1,\,e_2e_3b} + G_{[a_1...a_7|be_1,\,e_2e_3|a_8]}\right)\,\Lambda_{e_1e_2e_3}
$$
$$
-\,120\,\left(G_{a_1...a_8e_1e_2,\,e_3b} + G_{[a_1...a_7|be_1e_2,\,e_3|a_8]}\right)\,\Lambda_{e_1e_2e_3}
$$
$$
-\,110\,\left(G_{a_1...a_8e_1e_2e_3,\,b} + G_{[a_1...a_7|be_1e_2e_3,\,|a_8]}\right)\,\Lambda_{e_1e_2e_3}.
\eqno(2.13)$$
This last  result extends those given  in reference [8] by the addition of the level four fields. The extension of the $E_{11}$ algebra to include the level four commutators required to derive these additional results are given in appendix A. 
\par
As we have mentioned the above transformations apply when the Cartan forms are written as forms, that is, when written in the form 
$G_{ \underline \alpha}$ where $G_{ \underline \alpha}\equiv dz^\Pi G_{\Pi , \underline \alpha}$ and $G_{\Pi , \underline \alpha}$ are the components. 
However, the coordinates do transform under the rigid $g_0$ transformations and so therefore do the $G_{\Pi , \underline \alpha}$ on their $\Pi$ index. 
As a result we will use  the object  $G_{A , \underline \alpha} = (E^{-1})_A{}^\Pi G_{\Pi , \underline \alpha}$ which is inert under the rigid $E_{11}$ transformations,  but  transforms under the local $I_c(E_{11})$ transformations on all its indices.  One finds that these Cartan forms, when referred to the tangent space,  transform  on their first  index as [14]
$$
\delta G_{a, \bullet}= -3G^{b_1b_2}{}_{,\bullet}\ 
\Lambda_{b_1b_2 a},
\quad \delta G^{a_1a_2}{}_{, \bullet}= 6\Lambda^{a_1a_2
b}  G_{b,}{}_{\bullet}
\eqno(2.14)$$
These transformations are to be combined with the local transformations on the second $E_{11}$ index given earlier in this section. 
\par
The equations of motion are the set of equations that are invariant under the transformations of equations (2.10-2.13). To make progress we will work up to a given level in the coordinates and the fields. In this paper we will work up to and including the level four fields. However we will only work up to and including level one in the coordinates. When varying a given equation of motion we work so as to be sure to include all terms in the {\bf resulting equation of motion} that  contain derivatives with respect to the level zero  coordinates, that is, the usual coordinates of space time $x^\mu$. In other words we do not include the terms with derivatives with respect to level one coordinates in  the  equations resulting from the variation. 
Examining equation (2.14) we realise that in order to do this we must include all terms in the {\bf equation of motion we are  varying } that contain derivatives with respect to the level zero and one coordinates. 
As this is an important point we spell out the procedure in detail. We start with an equation of motion, generically denoted $E^{(n)}{}$, that only contains $n$ derivatives with respect to level zero coordinates and we vary it to find a new equation of motion that also contains only level zero  coordinates, however, in carrying out this step we find the terms in the original equation of motion that contain derivatives with respect to the level one coordinates, we  generically denoted the result by ${\cal E}^{(n)}$. In other words in this two step process we will first find $E^{(n)}{}$  and then   ${\cal E}^{(n)}$. 	Of course,  it would be better to carry out the variation more fully and  in one step but given the level of complexity we leave this to the   future. We will not write out $+\ldots$ in the equations of motion we will derive but take this to be understood in the sense just given. 

\par
\medskip 
{\bf 3. The three-six form duality relation }
\medskip
In our previous papers we have derived from the non-linear realisation  the unique $E_{11}$ invariant equation that is first order in derivatives and contains the  fields $A_{a_1a_2a_3}$ and $A_{a_1\ldots a_6}$, it was found to be given by  [14,8]
$$
{E^{(1)}}_{a_1\ldots a_4}\equiv { G}_{[a_1,a_2a_3a_4] }-{1\over 2.4!}\epsilon _{a_1a_2a_3a_4}{}^{b_1\ldots b_7} G_{b_1,b_2\ldots b_7 } =0
\eqno(3.1)$$
As explained at the end of section two at this stage we neglect any terms that contain derivatives with respect to level one coordinates. 
\par
From equation (3.1)  we can take two projections which result in  equations   that are second order in derivatives and  contain only one field. These equations are 
 given, at the linearised level,  by 
 
$$
E^{(2)}{}^{a_1a_2a_3}\equiv \partial_{b}E^{(1)}{}^{[b,a_1a_2a_3]} = \partial_{b}
{G }^{[b,a_1a_2a_3]} =0
\eqno(3.2)$$
and 
$$
{E^{(2)}}{}^{a_1\ldots a_6}\equiv {2\over 7!} \partial_{b}\epsilon ^{b  a_1\ldots a_6 c_1\ldots c_4 }E^{(1)}{}_{c_1\ldots c_4}=
\partial_{b}G^{[b , a_1\ldots a_6]}=0
\eqno(3.3)$$
These are of course the well known equations of motion for the three and six form fields. 

We will now vary these, two second order in derivatives, equations under the $I_c(E_{11})$ transformations of equations (2.10-2.13) beginning with equation   (3.2) to find that 

$$
\delta{\cal E}^{(2)}{}^{a_1a_2a_3 }= {3\over 2} E^{(2)}{}_b{}^{[a_1|}\Lambda ^{b | a_2a_3]} 
-{1\over 24} \epsilon ^{a_1 a_2 a_3 \nu\lambda_1\ldots \lambda_4 b_1b_2b_3}\partial_\nu  E^{(1)}{}_{\lambda_1\ldots \lambda_4} \Lambda_{ b_1 b_2 b_3}
$$
$$
= {3\over 2} E^{(2)}{}_b{}^{[a_1|}\Lambda ^{b | a_2a_3]} +3.5.7E^{(2)}{}^{a_1a_2a_3 b_1b_2b_3} \Lambda_{b_1b_2b_3}
 \eqno(3.4)$$
where 
$$
{\cal E}^{(2)}{}^{a_1a_2a_3 }=  E^{(2)}{}^{a_1a_2a_3 }
+ {1\over 4}\partial_b  G^{[a_1a_2 }{}^{,}{}^{ [  |b | a_3]] }
+{15\over 2} \partial_b G^{d_1d_2}{}_{, d_1d_2} {}^ {b a_1a_2a_3}
+{1\over 2}\partial^{[a_1 }\,G_{e}{}^{  a_2,( a_3 ] e  )} 
$$
$$
+{1\over 4}\partial_b  G^{[a_1a_2 }{}^{,}{}^{a_3 ] b } -{1\over 4}\partial^{[ a_1}  G^{a_2 a_3 ]}{}_{,}{}_{d}{}^{d}, 
\eqno(3.5)$$
$$
E^{(2)}{}_a{}^b\equiv R_a{}^b= \partial_a \omega_{c,}{}^{bc}-\partial _c \omega _{a, }{}^{bc} \ ,
\eqno(3.6)$$
and  
$$
\omega _{c, ab}= - G_{a, (bc)}+ G_{b, (ac)}+G_{c, [ab]}= - \partial_{a}h_{ (bc)}+ \partial_{b}h_{ (ac)}+ \partial_{c}h_{ [ab]}
\eqno(3.7)$$
which is the familiar expression for the spin connection; in the first and second equations we give the non-linear and linearized  expressions respectively. 
Thus varying the three  form equation of motion (3.2) we find the six  form equation of motion as well as a new equation $E^{(2)}{}_a{}^b=0$ which is just the linearised Einstein equation. 
\par
In line with the strategy spelt out at the end of section two we begin with the equations of motion (for example $E^{(2)}{}^{a_1a_2a_3}$) which contains only derivatives with respect to the level zero coordinates, that is,  the usual coordinates of space time  but we find the terms containing derivatives with respect to level one coordinates when we vary them (for example ${\cal E}^{(2)}{}^{a_1a_2a_3}$). The full non-linear versions of equations (3.2), (3.4), (3.5) and  (3.6) can be found in reference [7,8]. 
\par
The variation of the six form equation of motion (3.3) under the $I_c(E_{11})$ transformations of equation (2.10-2.13) is  given by 
$$
\delta {\cal E}^{(2)}{}_{a_1\ldots a_6}= {8\over 7}\Lambda_{[a_1a_2a_3}E^{(2)}{}_{a_4a_5a_6]} 
- 27.64 E^{(2)}{}_{a_1\ldots a_6 c_1c_2, c_3} \Lambda^{c_1c_2c_3}
\eqno(3.8)$$
where 
$$
{\cal E}^{(2)}{}_{a_1\ldots a_6}= { E^{(2)}{}^{a_1\ldots a_6}}-8\partial^d G^{c_1c_2}{}_{, d  a_1\ldots a_6 c_1, c_2} -36 \partial _{[d|} G^{c_1c_2}{}_{,| a_1\ldots a_6 c_1c_2],}{}^{d}
$$
$$
+{1\over 7} \partial_{[ a_1} G_{a_2a_3, a_4a_5a_6]}
\eqno(3.9)$$
and 
$$
E^{(2)}{}_{a_1\ldots a_8,}{}^{ b}\equiv 
-\,{1\over 4}\,\partial^{[d}\,G_{[d,\,a_1...a_8],\,}{}^{b]}
\eqno(3.10)$$
Thus varying the six  form equation of motion (3.3) we find the three form equation of motion as well as a new equation of motion, that is, $E^{(2)}{}_{a_1\ldots a_8,}{}^{ b}=0$. 
This equation of motion for the field $A_{a_1\ldots a_8,b}$ does correctly describe  gravity at the linearised level. One can verify, see appendix B,  that this equation of motion belongs to an irreducible representation of SL(11) which is consistent with the fact that $E_{11}$ only contains the  field $A_{a_1\ldots a_8,b}$ which has the same SL(11) irreducibility conditions. 
\par
Hence we have started from the three-six form duality relation of  equation (3.1) and derived the two equations which are second order in derivatives, that is, equations (3.2) and (3.3). We have then varied these two equations under the local $I_c(E_{11})$ transformations to find the  equations of motion 
$$
{ E^{(2)}{}}_{ab}= 0= { E^{(2)}{}}_{a_1\ldots a_8,}{}^{ b}
\eqno(3.11)$$
These   are the linearised equations of motion for gravity expressed in the standard  way using the usual metric and also expressed using the dual graviton field. 
We note that we do not have two physical gravitons as we will find in the next section that these two fields obey a first order duality relation.

\medskip 
{\bf 4. The gravity-dual gravity duality relation}
\medskip

We now consider the $E_{11}$ variation of the  three-six form  duality relation of equation (3.1). We will find a new duality relation that relates the usual field of gravity to the dual gravity field. We will then follow the same pattern as in the previous section; we will project this new duality relation   in two ways to find equations which are second order in derivatives, but only contain one field, and then we will compute the $E_{11}$ variations of these latter equations. 
\par
Under the $I_c(E_{11})$ transformations of equations (2.10-2.13) one finds that 
$$
\delta {\cal  E}^{(1)}{}_{a_1\ldots a_4}= {1\over 4!} \epsilon _{a_1\ldots a_4 }{}^{b_1\ldots b_7} \Lambda_{b_1 b_2b_3} E^{(1)}{}_{b_4 \ldots b_7}+ 
3\omega_{c ,}{}_{[a_1a_2} \Lambda ^c{}_{a_3a_4]}
$$
$$
-{7\over 2}  \epsilon_{a_1a_2a_3a_4}{}^{b_1\ldots b_7} G_{b_1, b_2\ldots b_7 c_1c_2,c_3} \Lambda ^{c_1c_2c_3} 
-{7\over 2}  \epsilon_{a_1a_2a_3a_4}{}^{b_1\ldots b_7} G_{c_1 , c_2 c_3 b_1 b_2\ldots b_6, b_7 } \Lambda ^{c_1c_2c_3} 
\eqno(4.1)$$
where 
$$
{\cal  E}^{(1)}{}_{a_1\ldots a_4}\equiv   {\cal G}_{a_1 a_2a_3a_4 }
-{1\over 2.4!}\epsilon _{a_1a_2a_3a_4}{}^{b_1\ldots b_7}{\cal  G}_{b_1 b_2\ldots b_6, b_7 }+{1\over 2} G_{[a_1a_2}{}_{,}{}_{a_3a_4]}
\eqno(4.2)$$
$$
{\cal G}_{a_1 a_2a_3a_4 }\equiv  G_{[a_1,a_2a_3a_4] }+{15\over 2}G^{b_1b_2}{}_{, b_1b_2 a_1\ldots a_4}
\eqno(4.3)$$
$$
{\cal G}_{a_1a_2 \ldots a_7 }\equiv G_{a_1,a_2 \ldots a_7 }
+28 G^{e_1e_2}{}_{, e_1e_2 [ b_1\ldots , b_7 ]}
\eqno(4.4)$$
\par
By substituting $\omega_{\lambda,\,\mu_1\mu_2} $ for $E^{(1)}{}_{\lambda,\,\mu_1\mu_2}$  one can show that  equation (4.1) can be written as 
$$
\delta {\cal   E}^{(1)}{}_{a_1\ldots a_4}= {1\over 4!} \epsilon _{a_1\ldots a_4 }{}^{b_1\ldots b_7} \Lambda_{b_1 b_2b_3}E^{(1)}{}_{b_4 \ldots b_7}+ 
3E^{(1)}{}_{c ,}{}_{[a_1a_2} \Lambda ^c{}_{a_3a_4]}
\eqno(4.5)$$
where 
$$
E^{(1)}{}_{\lambda,\,\mu_1\mu_2} \equiv \omega_{\lambda,\,\mu_1\mu_2} - {1\over 4}\,\varepsilon_{\mu_1\mu_2}{}^{\nu_1...\nu_9}\,G_{\nu_1,\,\nu_2...\nu_9,\,\lambda}
\eqno(4.6)$$

\par
This  non-trivial calculation is best carried out by first deducing the consequence of equation (4.1) by extracting $\Lambda^{c_1c_2c_3}$,   taking a double trace and using the fact that the variation of the first duality relation must vanish ($\delta {\cal  E}^{(1)}{}_{a_1\ldots a_4}=0$),  we   find  that 
$$
 E^{(1)}{}_{\lambda,}{}^{\,\mu_1\mu_2} -{4\over 7}E^{(1)}{}_{\rho,}{}^{\rho [\mu_1}\delta_\lambda ^{\mu_2]} \dot = 0
\eqno(4.7)$$
 In carrying out this step we have used  the identity 
$$3
X_{[a_1a_2}{}^{[c_1}\delta_{a_3a_4]}^{c_2c_3]}\delta_{c_2}^{a_3} \delta_{c_3}^{a_4} ={14\over 3} X_{a_1a_2}{}^{c_1}-{8\over 3} X_{d[a_1}{}^{d} \delta _{a_2]} ^{c_1}
\eqno(4.8)$$
for any tensor $X_{a_1a_2}{}^{c}$ which obeys $X_{a_1a_2}{}^{c}= X_{[a_1a_2]}{}^{c}$.  
As a result we conclude that    
$$
{ E}^{(1)}{}_{\lambda,\,\mu_1\mu_2} \equiv \omega_{\lambda,\,\mu_1\mu_2} - {1\over 4}\,\varepsilon_{\mu_1\mu_2}{}^{\nu_1...\nu_9}\,G_{\nu_1,\,\nu_2...\nu_9,\,\lambda}\dot = 0
\eqno(4.9)$$
The use of the symbol $\dot =$ will be discussed shortly 
\par 
Clearly, equation (4.9)  is a necessary condition for $\delta {\cal  E}^{(1)}{}_{a_1\ldots a_4}$to vanish but  looking at equation (4.5) we see  that it is also a sufficient condition.  That equation (4.1) can be rewritten in the form of equation (4.5) requires a  remarkable set of cancellations. Given that the equations we are deriving follow from the properties of the $E_{11}$ Dynkin diagram these cancellations  illustrate the magical  way $E_{11}$  leads to the correct dynamical equations. 
\par
Equation (4.9) is a duality relation between the usual gravity field and the dual gravity field. This relation was first proposed in reference [1] but it was found in the $E_{11}$ context in reference [14]. However, there are a number of subtle, but important,  features on how it should be interpreted. 
We recall that the  local Lorentz transformations were not used to fixed  our choice of group element of equation (2.1-2.2) and as such they  are still an explicit  symmetry. These transform the spin connection in the above equation by the inhomogeneous term,  $\delta \omega_{\lambda,\,\mu_1\mu_2} = \partial_\lambda \Lambda _{\mu_1\mu_2}+\ldots $ where $+\ldots $ indicate the homogeneous terms. As a result equation (4.9) is not invariant under local Lorentz transformations and we should consider it as being  valid only modulo local Lorentz transformations. In other words it is subject to the equivalence relation 
$$
E^{(1)}{}_{\lambda,\,\mu_1\mu_2}\sim E^{(1)}{}_{\lambda,\,\mu_1\mu_2} +\partial_{\lambda} \Lambda_{\mu_1\mu_2}+\ldots 
\eqno(4.10)$$
where $+\ldots $ indicate the homogeneous Lorentz transformations of $E^{(1)}{}_{\lambda,\,\mu_1\mu_2}$. This strategy was already advocated in reference [16]. The use of the symbol $\dot =$ implies that the equation only holds modulo the local transformations as just discussed. 
\par
In fact if one carries out the variation of $E^{(1)}{}_{a_1a_2a_3a_4} $ one finds not 
$\omega_{\lambda,\,\mu_1\mu_2}$ but the combination $\omega_{\lambda,\,\mu_1\mu_2} - \,G_{\lambda,\,[ \mu_1\mu_2 ]} $ provided one does not include the 
the last term in the definition of ${\cal E}^{(1)}{}_{a_1a_2a_3a_4} $ given in equation (4.2). On could choose not to include this term and then equation (4.9) would not hold  modulo local Lorentz transformations. that is, it would hold exactly. However,if the coefficients are not precisely as above one finds, carrying out the calculations later in this paper, that the duality relations do not close to form an $E_{11}$ invariant set of equations of motion. A closely related  point is that equation (4.5) holds exactly but  it contains $E^{(1)}{}_{c ,}{}_{[a_1a_2} $ which holds modulo the above transformations. However, precisely such a transformation is generated in this variation by the last term in equation (4.2) using equation (2.14).  
\par
As with the previous three-six form duality relation,  we can derive from equation (4.9) two second order equations that  each contain only one field. However, now we must do this in such a way as to find equations  that are invariant under the local Lorentz transformations. The equation that contains the usual graviton is found  by taking the  projection $\theta_1$ 
$$ 
E^{(2)}{}_\lambda{}^\mu = (\theta_1 E^{(1)}{})_\lambda{}^\mu \equiv  \partial_\nu E^{(1)}{}_{\lambda ,}{}^{\nu \mu}
-\partial_\lambda E^{(1)}{}_{\nu ,}{}^{\nu \mu}
= \partial_\nu \omega _{\lambda ,}{}^{\nu \mu} -\partial_\lambda \omega _{\nu ,}{}^{\nu \mu}=R_\lambda {}^\mu=0
\eqno(4.11)$$
where $R_\lambda {}^\mu$ is the Ricci tensor. We note that the dual graviton has dropped out and it is straightforward to show that equation (4.11)  is invariant under the local Lorentz transformations. We recall that we have already encountered  the symbol  $E^{(2)}{}_\lambda{}^\mu$ in equation (3.6). Thus from the gravity-dual gravity relation of equation (4.9) we have derive Einstein's equation at the linearised level along the lines given in reference [16]. 
\par
We now carry out another projection that is also invariant under the local Lorentz transformations, namely 
$$
E^{(2)}{}_{\nu_1\ldots \nu_8,}{}^{ \lambda}= (\theta_2 E^{(1)}{})_{\nu_1\ldots \nu_8,}{}^{ \lambda}\equiv 
-{1\over 2. 9!} \epsilon _{\nu_1\ldots \nu_8\tau\mu_1\mu_2 } \partial^{[\tau}E^{(1)}{}^{\lambda ],}{}^{\mu_1\mu_2}
$$
$$=
-\,{1\over 4}\,\partial^{[\tau}\,G_{[\tau,\,\nu_1...\nu_8],\,}{}^{\lambda]}=0
\eqno(4.12)$$

The reader can verify  that the usual gravity field drops out and one is left with the same second order equation of motion for the dual graviton that we encountered earlier in equation (3.11). That we recover our previous equations is to be expected as before, in section three,  we projected the three-six form equation (3.10), that is $E^{(1)}{}_{a_1\ldots a_4}$ to find second order equations which we then varied under $E_{11}$ transformation to find the higher level the second order equations (3.11), while in this section we have carried out the $E_{11}$ transformation of the three-six form equation $E^{(1)}{}_{a_1\ldots a_4}$ and then projected to find the same  two equations, (4.11) and (4.12).  
\par 
 We will now consider the $I_c(E_{11})$ variation of these last two equations. In fact the full non-linear variation of  $E^{(2)}{}_{a b}$ was found in reference [8] and for completeness we record the linearised result here 
$$
\delta {\cal E}^{(2)}{}_{ab} = -36 \Lambda ^{d_1d_2}{}_{ a} E^{(2)}{}_{bd_1d_2} -36 \Lambda ^{d_1d_2}{}_{ b} E^{(2)}{}_{ad_1d_2} +8\eta_{ab} E^{(2)}{}_{d_1d_2d_3}\Lambda ^{d_1d_2d_3}
\eqno(4.13)$$
where 
$$
{\cal E}^{(2)}{}_{ab}= {\cal  R}_{ab}- 6  \partial_{[b} G^{d_1d_2}{}_{, a d_1d_2]}\ ,
\eqno(4.14)$$ 
$$
{\cal  R}_{ab}= \partial_a \Omega_{c,}{}_{b}{}^{c}-\partial _c \Omega _{a, }{}_{b}{}^{c}
\eqno(4.15)$$
and 
$$
\Omega _{c, ab}\equiv \omega_{c, ab}
-3 G^{dc}{}_{, dab} -3 G^{d}{}_{b}{}_{, dac} +3 G^{d}{}_{a}{}_{, dbc} 
-\eta _{bc} G^{d_1d_2}{}_{, d_1d_2 a}+\eta _{ac} G^{d_1d_2}{}_{, d_1d_2 b}
\eqno(4.16)$$
\par

To find this result one can use that 
$$
\delta \Omega _{c, ab}=-18.2 \Lambda ^{ d_1d_2}{}_{ c} G_{[a,bd_1d_2]} -18.2 \Lambda ^{d_1d_2}{}_{ b} G_{[a,c d_1d_2 ]}-18.2 \Lambda ^{d_1d_2}{}_{ a} G_{[c,b d_1d_2]}
$$
$$
+8 \eta _{bc}  \Lambda ^{d_1d_2 d_3} G_{[a, d_1d_2d_3 ]}-8 \eta _{ac}  \Lambda ^{d_1d_2 d_3} G_{[b, d_1d_2d_3 ]}
\eqno(4.17)$$
The full non-linear versions of equations (4.13-17) can be found in reference [8]. 
\par
We  will now carry out  the variation of the   equation of motion (4.12) for the field $A_{a_1\ldots a_8,b}$, which is second order in the derivatives,  under  the $I_c(E_{11})$ transformations of equation (2.10-2.13). Our strategy for carrying the variation was spelt out at the end of section two and in the above discussions we have implemented it automatically by writing down the varied equation including its contributions that involve the level one coordinates. However, here we will carry out the derivation in two steps so that the reader can see how it works in detail. After quite some effort  we find that

$$
\delta E^{(2)}{}_{\,\rho_1...\rho_8,\,\lambda} =\delta \left(-\,{1\over 4}\,\partial_{[\nu}\,G_{[\nu,\,\rho_1...\rho_8],\,\lambda]}\right) = -\,{7\over 4}\,E^{(2)}{}_{\tau[\rho_1...\rho_5}\,
\Lambda_{\rho_6\rho_7 }{}^{\tau}\eta_{\rho_8]\lambda}\,
$$
$$
+\,275\,\left({\hat E^{(2)}{}}_{\nu\rho_1...\rho_8\sigma_1,\,\nu\sigma_2\sigma_3\lambda} - {1\over 9}\,{\hat E^{(2)}{}}_{\nu\rho_1...\rho_8\lambda,\,\nu\sigma_1\sigma_2\sigma_3}\right)\,\Lambda^{\sigma_1\sigma_2\sigma_3}
$$
$$
+\,{165\over 8}\,\left(\partial_{\nu}\,G_{[\nu,\,\rho_1...\rho_8\sigma_1\sigma_2],\,\sigma_3\lambda} - \partial_{\lambda}\,G_{[\nu,\,\rho_1...\rho_8\sigma_1\sigma_2],\,\sigma_3\nu} - {2\over 9}\,\partial_{\sigma_1}\,G_{[\sigma_2,\,\rho_1...\rho_8\lambda\nu],\,\sigma_3\nu} \right)\,\Lambda^{\sigma_1\sigma_2\sigma_3}
$$
$$
+\,{7\over 12}\,\partial_{\sigma_1}\,G_{\left[\sigma_2 , [\rho_1...\rho_6\right]}\,\delta_{\rho_7 , | \lambda}|\,\Lambda_{\rho_8 ]}{}^{\sigma_1\sigma_2}
$$
$$
-\,{55\over 8}\,\Big(20\,\partial_{\sigma_1}\,G_{[\nu,\,\rho_1...\rho_8\sigma_2],\,\sigma_3\lambda\nu} + {10\over 3}\,\partial_{\sigma_1}\,G_{[\nu,\,\rho_1...\rho_8\lambda],\,\sigma_2\sigma_3\nu}
$$
$$
+\,\partial_{\sigma_1}\,G_{\lambda,\,\rho_1...\rho_8\nu,\,\sigma_2\sigma_3\nu} - \partial_{\sigma_1}\,G_{\nu,\,\nu\rho_1...\rho_8,\,\sigma_2\sigma_3\lambda}\Big)\,\Lambda^{\sigma_1\sigma_2\sigma_3}
$$
$$
+\,{15\over 4}\,\left(\partial_{\sigma_1}\,G_{\nu,\,\nu\sigma_2\rho_1...\rho_8,\,\sigma_3\lambda} + {1\over 9}\,\partial_{\sigma_1}\,G_{\nu,\,\sigma_2\rho_1...\rho_8\lambda,\,\sigma_3\nu}\right)\,\Lambda^{\sigma_1\sigma_2\sigma_3}
$$
$$
-\,{10\over 3}\,\left(\partial_{\sigma_1}\,G_{[\rho_1,\,\rho_2...\rho_8]\lambda\sigma_2\nu,\,\sigma_3\nu} - \partial_{\sigma_1}\,G_{\lambda,\,\rho_1...\rho_8\sigma_2\nu,\,\sigma_3\nu}\right)\,\Lambda^{\sigma_1\sigma_2\sigma_3}
$$
$$
+\,{55\over 12}\,\Big(  \partial_{\sigma_1}\,G_{\nu,\,\sigma_2\sigma_3\nu\rho_1...\rho_8,\,\lambda} - \partial_{\sigma_1}\,G_{\nu,\,\sigma_2\sigma_3\nu\lambda[\rho_1...\rho_7,\,\rho_8]}
$$
$$
+\,{1\over 8}\,\partial_{\sigma_1}\,G_{\nu,\,\sigma_2\sigma_3\lambda\rho_1...\rho_8,\,\nu} - {9\over 8}\,\partial_{\sigma_1}\,G_{\lambda,\,\sigma_2\sigma_3\rho_1...\rho_8\nu,\,\nu}\Big)
\Lambda^{\sigma_1\sigma_2\sigma_3}.
\eqno(4.18)$$

To define the  other objects that appear in the above  equation we first define  
$$
E^{(2)}{}_{\rho_1...\rho_{10},\,\sigma_1...\sigma_4} \equiv \partial_{[\sigma_1}\,G_{[\rho_1,\,\rho_2...\rho_{10}],\,\sigma_2\sigma_3\sigma_4]} - {36\over 5\cdot 11!}\,\varepsilon_{\rho_1...\rho_{10}}{}^{\lambda}\,\partial_\lambda\,G_{[\sigma_1,\,\sigma_2\sigma_3\sigma_4]},
\eqno(4.19)$$
and then consider the quantity 
$$
{\hat E^{(2)}{}}_{\rho_1...\rho_{10},\,}{}^{\sigma_1...\sigma_4} \equiv E^{(2)}{}_{\rho_1...\rho_{10},\,}{}^{\sigma_1...\sigma_4} + {36\over 5\cdot 11!}\,\varepsilon_{\rho_1...\rho_{10}}{}^{\lambda}\,\partial_\lambda\,E^{(1)}{}^{\sigma_1...\sigma_4}
$$
$$
=\,\partial^{[\sigma_1}\,G_{[\rho_1,\,\rho_2...\rho_{10}],\,}{}^{\sigma_2\sigma_3\sigma_4]} + {3\over 55}\,\partial_\nu\,G_{[\rho_1,\,\rho_2...\rho_7}\,\delta_{\rho_8\rho_9\rho_{10}]\nu}^{\sigma_1\,\,\,.\,\,.\,\,.\,\,\,\sigma_4} + {21\over 220}\,\partial_\nu\,G_{\left[\nu,\,[\rho_1...\rho_6\right]}\,\delta_{\rho_7...\rho_{10}]}^{\sigma_1...\sigma_4},
\eqno(4.20)$$\
The effect of the second term in the middle  equation is to eliminate the Cartan form for the field $A_{a_1a_2a_3}$ and replace it by terms for the Cartan form for $A_{a_1\ldots a_6}$.
\par
Using equation (2.14) we can cancel all the terms of the generic form $G_{a, \bullet }\Lambda^{abc}\ldots $ by adding $-{1\over 6}G^{bc}{}_{,\bullet}\ldots $ and as a result we can write the equation (4.18) in the form 
$$
\delta{\cal E}^{(2)}{}_{\,\rho_1...\rho_8,\,\lambda} = -\,{7\over 4}\,E^{(2)}{}_{\sigma [\rho_1...\rho_5}\,\Lambda^{\sigma}{}_{ \rho_6\rho_7}\eta_{\rho_8] \lambda}\,
$$
$$
+\,275\,\left({\hat E^{(2)}{}}_{\nu\rho_1...\rho_8\sigma_1,\,\nu\sigma_2\sigma_3\lambda} - {1\over 9}\,{\hat E^{(2)}{}}_{\nu\rho_1...\rho_8\lambda,\,\nu\sigma_1\sigma_2\sigma_3}\right)\,\Lambda^{\sigma_1\sigma_2\sigma_3}
$$
$$
+\,{165\over 8}\,\left(\partial_{\nu}\,G_{[\nu,\,\rho_1...\rho_8\sigma_1\sigma_2],\,\sigma_3\lambda} - \partial_{\lambda}\,G_{[\nu,\,\rho_1...\rho_8\sigma_1\sigma_2],\,\sigma_3\nu} - {2\over 9}\,\partial_{\sigma_1}\,G_{[\sigma_2,\,\rho_1...\rho_8\lambda\nu],\,\sigma_3\nu}\right)\,\Lambda^{\sigma_1\sigma_2\sigma_3}
\eqno(4.21)$$
where 
$$
{\cal E}^{(2)}{}_{\,\rho_1...\rho_8,\,\lambda}= { E^{(2)}{}}_{\,\rho_1...\rho_8,\,\lambda}
-\,{7\over 6.12}\,\partial_{[\sigma |}\,G^{ \sigma }{}_{[\rho_1, |\rho_2...\rho_7 ]}\,\eta_{\rho_8]\lambda}
$$
$$
-\,{55\over 48}\,\Big(20\,\partial_{[\nu |}\,G^{\sigma_2\sigma_3}{}_{,|\,\rho_1...\rho_8\sigma_2],\,\sigma_3\lambda}{}^{\nu} + {10\over 3}\,\partial_{[\nu |}\,G^{ \sigma_2\sigma_3}{}_{,\,|\rho_1...\rho_8\lambda],\,\sigma_2\sigma_3}{}^{\nu}
$$
$$
+\,\partial_{\lambda }\,G^{ \sigma_2\sigma_3}{}_{,\,\rho_1...\rho_8\nu,\,\sigma_2\sigma_3}{}^{\nu} - \partial^\nu\,G^{ \sigma_2\sigma_3}{}_{,\,\nu\rho_1...\rho_8,\,\sigma_2\sigma_3\lambda}\Big)\,
$$
$$
+\,{5\over 8}\,\left(\partial^{\nu}\,G^{ \sigma_2\sigma_3}{}_{,\,\nu\sigma_2\rho_1...\rho_8,\,\sigma_3\lambda} + {1\over 9}\,\partial^{\nu}\,G^{ \sigma_2\sigma_3}{}_{,\,\sigma_2\rho_1...\rho_8\lambda,\,\sigma_3\nu}\right)\,
$$
$$
-\,{10\over 3.6}\,\left(\partial_{[\rho_1 |}\,G^{ \sigma_2\sigma_3}{}_{,\,|\rho_2...\rho_8]\lambda\sigma_2\nu,\,\sigma_3}{}^{\nu} - \partial_{\lambda}\,G^{ \sigma_2\sigma_3}{}_{,\,\rho_1...\rho_8\sigma_2\nu,\,\sigma_3}{}^{\nu}\right)\,
$$
$$
+\,{55\over 72}\,\Big(\partial^{\nu}\,G^{ \sigma_2\sigma_3}{}_{,\,\sigma_2\sigma_3\nu\rho_1...\rho_8,\,\lambda} - \partial^{\nu}\,G^{ \sigma_2\sigma_3}{}_{,\,\sigma_2\sigma_3\nu\lambda[\rho_1...\rho_7,\,\rho_8]}
$$
$$
+\,{1\over 8}\,\partial^\nu\,G^{ \sigma_2\sigma_3}{}_{,\,\sigma_2\sigma_3\lambda\rho_1...\rho_8,\,\nu} - {9\over 8}\,\partial_{\lambda}\,G^{ \sigma_2\sigma_3}{}_{,\,\sigma_2\sigma_3\rho_1...\rho_8\nu,\,}{}^{\nu}\Big).
\eqno(4.22)$$

The reader may notice that the last term of equation (4.21) could also be removed  in the same  way. However, we must keep this term as it ensures that the right-hand side of the equation possess the same   SL(11) irreducibility properties as the left-hand side. 
\par
Since ${\cal E}^{(2)}{}_{\,\rho_1...\rho_8,\,\lambda}=0$ we conclude that the right-hand side of equation (4.20) vanishes and extracting off $\Lambda ^{\sigma_1\sigma_2\sigma_3}$,  using the previously derived equations of motion 
$E^{(2)}{}_{\sigma_1...\sigma_6} = 0 $, we find the equations 
$$
E^{(2)}{}_{\rho_1...\rho_9,\,}{}^{\sigma_1\sigma_2\sigma_3} \equiv E^{(2)}{}_{\nu\rho_1...\rho_9,\,}{}^{\nu\sigma_1\sigma_2\sigma_3} 
$$
$$
=\partial^{[\nu}\,G_{[\nu,\,\rho_1...\rho_9],}{}^{\sigma_1\sigma_2\sigma_3]} - {36\over 5\cdot 11!}\,\varepsilon_{\rho_1...\rho_9}{}_{\lambda_1\lambda_2}\,\partial^{\lambda_1}\,G^{[\lambda_2,\,\sigma_1\sigma_2\sigma_3]} = 0
\eqno(4.23)$$
and $$
E^{(2)}{}_{\nu_1  \nu_2\ldots \nu_{11} ,   \kappa\tau, \rho}  \equiv \partial _\tau G_{[\nu_1 , \nu_2\ldots \nu_{11}] ,  \rho \kappa} - \partial _\kappa G_{[\nu_1 , \nu_2\ldots \nu_{11} ],  \rho \tau} \dot = 0.
\eqno(4.24)$$
The appearance of the symbol $\dot =$ in this equation is the subject of section six where this last equation for the field $A_{a_1\ldots a_{10}, b_1b_2}$ is analysed.  
We note that 
$$
\hat E^{(2)}{}_{\rho_1...\rho_9,\,}{}^{\sigma_1\sigma_2\sigma_3} \equiv 
 E^{(2)}{}_{\rho_1...\rho_9,\,}{}^{\sigma_1\sigma_2\sigma_3} + {36\over 5\cdot 11!}\,\varepsilon_{\rho_1...\rho_9}{}_{\lambda_1\lambda_2}\,\partial^{\lambda_1}\,E^{(1)}{}^{[\lambda_2,\,\sigma_1\sigma_2\sigma_3]}
\eqno(4.25)$$
and so $\hat E^{(2)}{}_{\rho_1...\rho_9,\,}{}^{\sigma_1\sigma_2\sigma_3} = E^{(2)}{}_{\rho_1...\rho_9,\,}{}^{\sigma_1\sigma_2\sigma_3} $ when we use the three-six form duality equation. 
\par
Equations (4.23) and (4.24) are the necessary and sufficient conditions for the right hand side of equation (4.21) to vanish  as it  can be rewritten  as 
$$
\delta{\cal E}^{(2)}{}_{\,\rho_1...\rho_8,\,\lambda} = -\,{7\over 4}\,E^{(2)}{}_{\sigma [\rho_1...\rho_5 }\,\Lambda^{\sigma }{}_{\rho_6\rho_7}\eta_{\rho_8]\lambda}\,
$$
$$
+\,275\,\left({\hat E^{(2)}{}}_{\rho_1...\rho_8\sigma_1,\,\sigma_2\sigma_3\lambda} - {1\over 9}\,{\hat E^{(2)}{}}_{\rho_1...\rho_8\lambda,\,\sigma_1\sigma_2\sigma_3}\right)\,\Lambda^{\sigma_1\sigma_2\sigma_3}
$$
$$
+\,{165\over 8}\,\left(E^{(2)}{}_{\nu \rho_1...\rho_8\sigma_1\sigma_2, \nu\lambda , \sigma_3}  - {1\over 9}\,E^{(2)}{}_{\sigma_2\rho_1...\rho_8\lambda\nu, \sigma_1\sigma_3 , \nu}\right)\,\Lambda^{\sigma_1\sigma_2\sigma_3}
\eqno(4.26)$$
\par
We now discuss the equations of motion of  equation (4.23)  for the  level four field $A_{a_1\ldots a_9, b_1b_2b_3}$  that appear in the dynamics. We note that equation (4.23) is unlike  the previous equations of motion that were second order in derivatives that we have found in that it involves two fields rather than  a single field. However, as we will comment on later in the paper,  in contrast to the duality relations that are first order in derivatives this equation is gauge invariant. 
In order to eliminate the field $A_{a_1a_2a_3}$ in   equation (4.23) we must take the triple trace   to find the equation 
$$
E^{(2)}{}_{\rho_1...\rho_6 \nu_1\ldots \nu_4,}{}^{\nu_1\dots \nu_4} = 
\partial^{[\nu_1} G_{[\rho_1, ..\rho_6 \nu_1\ldots \nu_4 ],}{}^{\nu_2\dots \nu_4]}=0
\eqno(4.27)$$
This is indeed the correct equation of motion for the $A_{a_1\ldots a_9, b_1b_2b_3}$ to describe the same degrees of freedom which are usually encoded in the three form [11]. 
\par
 We note that there is no equation of motion for the field 
$A_{a_1\ldots a_{11}, b}$ as the terms that involve this field  can be removed by adding terms to 
${\cal E}^{(2)}{}_{\,\rho_1...\rho_8,\,\lambda}$ as we have indeed done by adding the  last terms in equation (4.22).

\medskip 
{\bf 5. Variation of the gravity-dual gravity relation }
\medskip
In the last section we  carried out the $I_c(E_{11})$ variation of the three-six form duality relation of equation (3.1) to find the gravity-dual gravity duality relation of equation (4.9). In this  section we will carry out the $I_c(E_{11})$ variation of this  gravity-dual gravity  relation of equation (4.9) to find a new duality relation. Following our strategy, as outlined at the end of section two, we find that 
$$
\delta{\cal E}^{(1)}{}_{\lambda,\,\mu_1\mu_2} = {7\over 12}\,\varepsilon_{\mu_1\mu_2}{}^{\nu_1...\nu_6\sigma_1\sigma_2\sigma_3}\, E^{(1)}{}_{\lambda\nu_1...\nu_6}\,\Lambda_{\sigma_1\sigma_2\sigma_3} + {1\over 2}\,\varepsilon_{\mu_1\mu_2}{}^{\nu_1...\nu_7\sigma_1\sigma_2}\, E^{(1)}{}_{\nu_1...\nu_7}\,\Lambda_{\sigma_1\sigma_2\lambda}
$$
$$
+{55\over 2} \Lambda_{\sigma_1\sigma_2 [\mu_1}\epsilon _{\mu_2 ] }{}^{\nu_1\ldots \nu_{10} } E^{(1)}{}_{\nu_1\ldots \nu_{10} , \lambda}{}^{\sigma_1\sigma_2}
-{55\over 18}  \Lambda^{\sigma_1\sigma_2 \sigma_3}\eta_{\lambda[\mu_1} \epsilon _{\mu_2]}{}^{ \nu_1\ldots \nu_{10} } E^{(1)}{}_{\nu_1\ldots \nu_{10} , \sigma_1\sigma_2\sigma_3}  
$$
$$
+ {3\over 4}\,\Lambda_{\mu_1\mu_2}{}^{\sigma}\,\varepsilon^{\rho_1...\rho_{11}}\,G_{\rho_1,\,\rho_2\ldots \rho_{11},\sigma\lambda}
+\partial_\lambda \tilde \Lambda_{\mu_1\mu_2} 
\eqno(5.1)$$
where 
$$
{\cal E}^{(1)}{}_{\lambda,\,\mu_1\mu_2} = \Omega_{\lambda,\,\mu_1\mu_2}  - {1\over 4}\,\varepsilon_{\mu_1\mu_2}{}^{\nu_1...\nu_9}\,G_{\nu_1,\,\nu_2...\nu_9,\,\lambda}
$$
$$
-\,\varepsilon_{\mu_1\mu_2}{}^{\nu_1...\nu_9}\,\Bigg[{55\over 3}\,\left({1\over 9}\,G^{\sigma_1\sigma_2 }{}_{, \nu_1...\nu_9,\,\sigma_1\sigma_2\lambda} + {1\over 8}\,G^{\sigma_1\sigma_2}{}_{, \nu_1...\nu_8\lambda,\,\sigma_1\sigma_1\nu_9}\right)
$$
$$
+\,10\,\left({1\over 9}\,G^{\sigma_1\sigma_2}{}_{,\nu_1...\nu_9\sigma_1,\,\sigma_2\lambda} + {1\over 8}\,G^{\sigma_1\sigma_2}{}_{, \nu_1...\nu_8\lambda\sigma_1,\,\sigma_2\nu_9}\right)
$$
$$
+\,{55\over 4}\,\left({1\over 9}\,G^{\sigma_1\sigma_2}{}_{,\sigma_1\sigma_2\nu_1...\nu_9,\,\lambda} + {1\over 8}\,G^{\sigma_1\sigma_2}{}_{,\sigma_1\sigma_2\lambda\nu_1...,\,\nu_9}\right)\Bigg] , 
\eqno(5.2)$$  

$$
 E^{(1)}{}_{\mu_1\ldots \mu_{10} , \sigma_1\sigma_2\sigma_3}
\equiv G_{[\mu_1 , \ldots \mu_{10}] , \sigma_1\sigma_2\sigma_3}
-{1\over 5.5.11.7!} \epsilon _{\mu_1\ldots \mu_{10}}{}^{\tau} G_{[\tau,  \sigma_1\sigma_2\sigma_3] }
\eqno(5.3)$$
$$
E^{(1)}{}_{\sigma_1...\sigma_7} \equiv {2\over 7!}\epsilon_{\sigma_1...\sigma_7} {}^{\rho_1\ldots \rho_4} E^{(1)}{}_{ \rho_1\ldots \rho_4}=  G_{[\sigma_1,\,...\sigma_7]} + {2\over 7!}\,\varepsilon_{\sigma_1...\sigma_7}{}^{\nu_1...\nu_4}\,G_{\nu_1,\,\nu_2\nu_3\nu_4}
\eqno(5.4)$$
and 
$$
\partial_\lambda \tilde \Lambda_{\mu_1\mu_2}= 
-\,\varepsilon_{\mu_1\mu_2}{}^{\nu_1...\nu_9}\,\,\Bigg[{1\over 12}\,G_{\lambda,\,\nu_1...\nu_6}\,\Lambda_{\nu_7\nu_8\nu_9} + {55\over 36}\,G_{\lambda,\,\nu_1...\nu_9,\,\sigma_1\sigma_2\sigma_3}\Lambda^{\sigma_1\sigma_2\sigma_3}
$$
$$
+\,{15\over 4}\,G_{\lambda,\,\nu_1...\nu_8\sigma_1\sigma_2,\,\sigma_3\nu_9}
\Lambda^{\sigma_1\sigma_2\sigma_3} + {55\over 16}\,G_{\lambda,\,\sigma_1\sigma_2\sigma_3\nu_1...,\,\nu_9}\Lambda^{\sigma_1\sigma_2\sigma_3}\Bigg],
\eqno(5.5)$$
It will be useful   to introduce the quantity  $\hat  E^{(1)}{}_{\tau , \sigma_1 \sigma_2 \sigma_3}$ using the equation 
$$
 E^{(1)}{}_{\nu_1 \nu_2...\nu_{10},\,\sigma_1\sigma_2\sigma_3}= G_{[\nu_1,\,\nu_2...\nu_{10} ],\,\sigma_1\sigma_2\sigma_3}
- {1\over 25.11.7!}\epsilon _{\nu_1 \nu_2...\nu_{10}}{}^{\tau} G_{[\tau , \sigma_1 \sigma_2 \sigma_3]}
$$
$$
\equiv  - {1\over 25.11.7!}\epsilon _{\nu_1 \nu_2...\nu_{10}}{}^{\tau}\hat  E^{(1)}{}_{\tau , \sigma_1 \sigma_2 \sigma_3}
$$
This object can  also be expressed as 
$$
\hat  E^{(1)}{}_{\tau, \sigma_1 \sigma_2\sigma_3} \equiv  G_{[\tau,\sigma_1\sigma_2\sigma_3]} + {55\over 144}\,\varepsilon_{\tau}{}^{\nu_1...\nu_{10}}\,G_{\nu_1,\,\nu_2...\nu_{10},\,\sigma_1\sigma_2\sigma_3}
$$
$$= {55\over 144} \epsilon _{\tau}{}^ {\nu_1\ldots \nu_{10} }  E^{(1)}{}_{\nu_1,\nu_2\ldots \nu_{10}, \sigma_1 \sigma_2\sigma_3}
\eqno(5.6)$$
\par
We  now discuss the consequences that can be drawn from  equation (5.1).  As we have previously derived the equation  $ E^{(1)}{}_{\nu_1...\nu_7}=0$,  we can set the first two terms on the right-hand side to zero. We recall from equation (4.10) that ${\cal E}^{(2)}{}_{\lambda,\,\mu_1\mu_2}$ only vanishes modulo local Lorentz transformations and so we must take this into account when we vary it, as a result  we can discard term in the variation that are of this form, that is the last term in equation (5.1). The net result is that the right-hand side of equation (5.1) becomes 
$$
\delta{\cal E}^{(2)}{}_{\lambda,\,\mu_1\mu_2} ={55\over 2} \Lambda_{\sigma_1\sigma_2 [\mu_1}\epsilon _{\mu_2 ] }{}^{\nu_1\ldots \nu_{10} } E^{(1)}{}_{\nu_1\ldots \nu_{10} , \lambda}{}^{\sigma_1\sigma_2}
$$
$$
-{55\over 18}  \Lambda^{\sigma_1\sigma_2 \sigma_3}\eta_{\lambda[\mu_1} \epsilon _{\mu_2]}{}^{ \nu_1\ldots \nu_{10} } E^{(1)}{}_{\nu_1\ldots \nu_{10} , \sigma_1\sigma_2\sigma_3}  
+ {3\over 4}\,\Lambda_{\mu_1\mu_2}{}^{\sigma}\,\varepsilon^{\rho_1...\rho_{11}}\,G_{\rho_1,\,\rho_2\ldots \rho_{11},\sigma\lambda} 
\eqno(5.7)$$
Extracting off the arbitrary $\Lambda^{\sigma_1\sigma_2 \sigma_3}$ and taking a double trace we conclude that 
$$
 E^{(1)}{}_{\nu_1\ldots \nu_{10} , \sigma_1\sigma_2\sigma_3}  \dot = 0 
\eqno(5.8)$$
and 
$$
 E^{(1)}{}_{\rho_1\,\rho_2\ldots \rho_{11},\sigma\lambda}\equiv G_{[\rho_1,\,\rho_2\ldots \rho_{11}],\sigma\lambda}\dot = 0 
\eqno(5.9)$$
These two equations only hold modulo certain transformations,  hence the use of the symbol $\dot =$. 
\par
We have encountered four  duality relations, the three-six form duality equation (3.1), the gravity-dual gravity equation (4.9) and now equations (5.8) and (5.9).  The first of these held as an equation in the usual way, however,  the gravity-dual gravity duality was found  to vanish only modulo  local Lorentz transformations. From equations (3.1) and (4.9)  we have taken projections to find equations that are second order in derivatives the first of which was the familiar equation of motion for the three form.  Starting from this latter  equation of motion (3.2) and 
 by taking repeated  $E_{11}$ variations we found the equations of motion for the six form (3.3)  and  graviton  (3.11), the dual graviton  (3.11),  equation (4.23) for the field $A_{a_1\ldots a_9, b_1b_2b_3}$. All these equations hold without any modulo considerations. The status of  equation (4.24) for the field  $A_{a_1\ldots a_{10} , b_1b_2}$ is the subject of section six. These steps are summarised in table one in section one. 
\par
We will now discuss how to integrate the equations which are second order in derivatives to find the duality relations  that are first order in derivatives. In this way we will  find in which sense the duality relations hold. We begin with the very familiar equation of motion (3.2) for the three form, $\partial_\nu G^{[\nu , \mu_1\mu_2\mu_3]}= 0$ which,  by 
defining the quantity, 
$F_{\tau_1\ldots \tau_7}$ by 
$$
{1\over 48}  \epsilon _{ \nu_1\ldots \nu_4}{}^{ \tau_1\ldots \tau_7} F_{\tau_1\ldots \tau_7}\equiv G_{[\nu_1 , \ldots \nu_4 ]}
\eqno(5.10)$$ 
we can write in the form 
$$
\partial_\nu G^{[\nu , \mu_1\mu_2\mu_3]}= {1\over 48}  \epsilon ^{\nu \mu_1\mu_2 \mu_3 \tau_1\ldots \tau_7} \partial_\nu
F_{\tau_1\ldots \tau_7}=0
\eqno(5.11)$$
 This equation implies that $F_{\tau_1\tau_2\ldots \tau_7}= \partial_{[\tau_1}A_{\tau_2\ldots \tau_7]}$. Substituting this result into equation (5.10)  we recover an equation that is first order in derivatives and is the three-six form duality relation of equation (3.1).  Indeed one can run a similar argument for the six form equation of motion (3.3) to find the three form and the same duality relation. Of course we identify the six form and three forms we find in this process with the fields we already have. Although we have stripped off one derivative we do not find that the resulting equation has any additional symmetries as the solution to equation (5.11) just involves the field $A_{\tau_1\ldots \tau_7}$. 
\par
We can repeat this analysis for the graviton equation $R_{ab}=0$ and in the same way one finds the gravity-dual gravity equation (4.9) which also involves the field $A_{a_1\ldots a_8,b}$. A result one can also find by integrating the equation of motion (4.12) for the field $A_{a_1\ldots a_8,b}$. However, in this case we find that the resulting duality relation  only holds modulo local Lorentz transformations. 
\par
Next  we turn our attention to the  field $A_{a_1\ldots a_9, b_1b_2b_3}$ whose equation of motion was given in equation (4.27). However, this was a consequence of equation (4.23) which is a relation between the fields strengths of the three form field and the $A_{a_1\ldots a_9, b_1b_2b_3}$ field. One can integrate, in the above sense, either of these equations 
and find a duality relation, namely the relation that $ E^{(1)}{}_{\mu_1\ldots \mu_{10} , \sigma_1\sigma_2\sigma_3}$ vanishes. This was in effect carried out in reference [11] and in particular in equations (3.2.7) onwards in this reference. It was found that one arrives at the relation 
$$
 E^{(1)}{}_{\mu_1\ldots \mu_{10} ,}{}^{ \sigma_1\sigma_2\sigma_3}
$$
$$=G_{[\mu_1 , \ldots \mu_{10}] ,}{}^{ \sigma_1\sigma_2\sigma_3}
-{1\over 5.5.11.7!} \epsilon _{\mu_1\ldots \mu_{10}}{}_{\tau} G^{[\tau,  \sigma_1\sigma_2\sigma_3] }
+ \partial^{[\sigma_1}\partial_{[\mu_1}\Lambda_{\ldots \mu_{10}] ,}{}^{ \sigma_2\sigma_3 ]}
\eqno(5.12)$$
In other words the relation is modulo the transformation with parameter 
$\Lambda_{\mu_1\ldots \mu_{10} ,}{}^{ \sigma_2\sigma_3 }$. We recognise this is precisely the duality relation of equation (5.8) but we now realise that it holds only in the way just described. In reference [11] it is also explained how to start with the equation of motion for the three form (3.2) and find the duality relation just mentioned. 
\par
Although the above discussion on the transformations that the duality  relations are modulo has been  derived entirely from an  $E_{11}$ viewpoint, it would be good to find a more systematic and efficient  derivation. As a step in this direction we now consider the matter  from a different viewpoint. We observe that the equations of motion which are second order in derivatives that we have found for the three form, six form, graviton dual graviton and  the $A_{a_1\ldots a_9, b_1b_2b_3}$ field are invariant under the gauge  transformations 
$$
\delta A_{a_1a_2a_3}= \partial_{[ a_1}\Lambda_{a_2a_3]} ,\ \delta A_{a_1\dots a_6}= \partial_{[ a_1}\Lambda_{a_2\ldots a_6]} ,\ 
\delta h_{ab}= \partial_a \xi_b +\partial_b \xi_a ,\ 
$$
$$
\delta h_{a_1\ldots a_8, b} = \partial_{[a_1} \Lambda_{a_2\ldots a_8],b}
+ \partial_{b} \hat \Lambda_{ a_1 a_2\ldots a_8}-\partial_{[a_1} \hat \Lambda_{a_2\ldots a_8]b}
$$
$$
\delta A_{a_1\ldots a_9, b_1b_2b_3}= 9\partial_{[a_1}\Lambda^{(1)}{}_{a_2\ldots a_9], b_1b_2b_3}+
(\partial_{[b_1 |}\Lambda^{(2)}{}_{a_1a_2\ldots a_9], |b_2b_3]}+{9\over 7}\partial_{[a_1}\Lambda^{(2)}{}_{a_2\ldots a_9][b_1, b_2b_3]}
)
\eqno(5.13)$$
The last transformation was given in reference [11] and the reader can verify that it leaves the quantity of equation (4.19) invariant and so also equation (4.23).  In fact these are the expected gauge transformations for these fields and they are the gauge transformations that were derived from the $E_{11}$ viewpoint in reference [18]. 
\par
Although the  three-six form duality relation of equation (3.1) is gauge invariant, like the second order field equations considered above, the  other equations that are first order in derivatives that we have derived are not. In particular the gravity-dual gravity relation of equation (4.9) transforms under the above gauge transformations  of equation (5.13) as 
$$
\delta  E^{(1)}{}_{\lambda,\,\mu \nu} = \partial_\lambda (-\partial_\mu  \xi_\nu+\partial_\nu  \xi_\mu -{1\over 4} \epsilon_{\mu\nu}{}^{\tau_1\ldots \tau_9}\partial_{\tau_1} \hat \Lambda_{\tau_2\ldots \tau_9} )
\eqno(5.14)$$ 
which we recognise as a local Lorentz transformation.  However, as this duality  relation only holds modulo local Lorentz transformations it is invariant but only once we take these transformations into account. 
The same holds of the duality relation of equation (5.8),  carrying out the gauge transformation of equation (5.13) we find that it gives 
gives precisely the terms which this  equation holds modulo,  as listed in equation (5.12). Hence we expect all that the equations derived from $E_{11}$  which are first order in derivatives, with the exception of the three-six form duality relation,  will only hold modulo certain transformations and these transformations will be associated with gauge transformations. 
\par
We can also view things  in the inverse way. Given the duality relations, including the transformations that they hold modulo,  we can derive the equations of motion as we did for the fields in the gravity sector in section four. We note that from the duality equation (5.8), or equivalently (5.12)  we can derive an equation that is second order in derivatives and is gauge invariant. The simplest equation one derives in this way is 
$$
E^{(2)}{}_{\rho_1...\rho_{10},\,\sigma_1...\sigma_4} \equiv \partial_{[\sigma_1}\,G_{[\rho_1,\,\rho_2...\rho_{10}],\,\sigma_2\sigma_3\sigma_4]} - {36\over 5\cdot 11!}\,\varepsilon_{\rho_1...\rho_{10}}{}^{\lambda}\,\partial_\lambda\,G_{[\sigma_1,\,\sigma_2\sigma_3\sigma_4]}=0
\eqno(5.15)$$
in other words we find the quantity defined in  equation (4.19). It is the single trace of this that we previously found to be an equation of motion, that is, equation (4.23). From equation (5.15) it is straightforward to derive the three form equation of motion (3.2) and the equation of motion of equation (4.26) for the $A_{a_1\ldots a_{9}, b_1b_2b_3}$.


\medskip
{\bf 6. The equation of motion for the field $A_{a_1\ldots a_{10}, b_1b_2}$ }
\medskip
In section four we carried out the $E_{11}$ variation of  the equation of motion for the dual graviton field $A_{a_1\ldots a_{8}, b}$ and we found  that the field 
$A_{a_1\ldots a_{10}, b_1b_2}$ obeyed equation (4.24). This equation has two derivatives. However, the field $A_{a_1\ldots a_{10}, b_1b_2}$ has in effect three blocks of antisymmetrised indices, the symmetrised pair count as two blocks, and so one would normally expect its equation of motion to have three rather than two derivatives, that is, one derivative for each block of indices. This raises the suspicion that equation (4.24) may only hold modulo certain transformations. We recall that the equations derived previously that were first order in derivatives, generically denoted $E^{(1)}$, only held modulo certain transformations. The one exception being of the very first one. However, apart from the equation (4.24) for the field $A_{a_1\ldots a_{10}, b_1b_2}$ all the equations which were  second order in derivatives, generically denoted $E^{(2)}$,  and were derived from the $E^{(1)}$ equations and whose $E_{11}$ transformations lead to existing equations did hold exactly. We also noted that all these equations were gauge invariant and did satisfy the rule that their equations of motion had the same number of derivatives as the corresponding field had blocks of indices. 
\par
 Certainly as we consider fields at higher and higher levels they will possess more and more blocks of indices and it must be the case that their   field equations, that is, the equations that hold exactly, must   have higher and higher number of derivatives. Hence as we go to higher levels there must come a point at which the equations of second order in derivatives only hold modulo certain transformations. To test which possibility holds for equation (4.24) 
we can take its $E_{11}$ variation and see how it varies into our previously derived   equations. If it varies into the previous equations without having to take account of any modulo transformations then we can conclude that it will hold exactly but  if it requires such transformations it can not hold exactly. 
However as a first, and simpler step,  we will examine if equation (4.24)  is gauge  invariant. 
\par
The gauge transformations of the $E_{11}$ fields were found in reference [18].  Using equation (5.12) of this reference we find that the gauge transformations of the field $A_{a_1\ldots a_{10},\,b_1b_2}$ are  given by    
$$
\delta A_{a_1\ldots a_{10},\,b_1b_2} = \left( D_{a_1...a_{10},\,b_1b_2}\right)_ D{}^C\,\partial_C\,\Lambda^D.
\eqno(6.1)$$
Here $\Lambda^D$ is the parameter of the gauge transformation. In this equation  $\partial_C$ are the derivatives with respect to the generalised coordinates and the parameters $\Lambda^D$ are labelled by the $l_1$ representation.  The matrix  $ D_{a_1...a_{10},\,b_1b_2}$ is the one that occurs in the commutator of the generator $R^{a_1...a_{10},\,b_1b_2}$ and the generator  $l_D$ in the $l_1$ representation to give the generator $l_C$. We are interested in the terms in the gauge transformations that contain the usual spacetime derivatives, that is, those of level zero. As a result we require a parameter $\Lambda^D$ in equation  (6.1) that  belongs to level four, like the field we are varying. By examining the commutators one finds that  only the last three $l_1$ generators given in  equation (A.5) in appendix A can contribute.  As already mentioned each of these three generators has a gauge parameter associated with it, namely:
$$
\Lambda_{a_1...a_9,\,b_1b_2},\quad \Lambda_{a_1...a_{10},\,b}^{\left(1\right)},\quad \Lambda_{a_1...a_{10},\,b}^{\left(2\right)}.\eqno(6.2)
$$
We note that two of the parameters have the same index structure corresponding to the fact that the corresponding $l_1$ generators  occur with multiplicity two in the $l_1$ representation. In fact  these two parameters contribute to the right-hand side of (6.1) in  the same way  and so we may just take one of them.  These leaves us with only two parameters: $\Lambda_{a_1...a_9,\,b_1b_2}$ and $\Lambda_{a_1...a_{10},\,b}$. Using equation (6.1) and equations (A.6-A.8) in appendix A we find that 
$$
\delta A_{a_1\ldots a_{10},\,b_1b_2} = \partial_{(b_1}\,\Lambda_{|a_1...a_{10}|,\,b_2)} - {10\over 11}\,\partial_{[a_1}\,\Lambda_{a_2...a_{10}]( b_1,\,b_2 )}
$$
$$
+\,\partial_{[a_1}\,\Lambda_{a_2...a_{10}],\,b_1b_2} 
\eqno(6.3)$$
In deriving this last equation  we have rescaled the $\Lambda$ parameters in order to get simpler coefficients.
\par
Equation (4.24) involves the Cartan form, corresponding to the field 
$A_{|a_1...a_{10}|,b_1b_2)}$,  which is antisymmetrised in its first eleven indices and this transforms as 
$$
\delta G_{[a_1, a_2\ldots a_{11}], b_1b_2}= {1\over 2}(\partial_{b_1}\,\partial_{[a_1}\Lambda_{ a_2...a_{11}],\,b_2}+\partial_{b_2}\,\partial_{[a_1}\Lambda_{ a_2...a_{11}],\,b_1})
\eqno(6.4)$$
\par
We see that $\epsilon ^{a_1 a_2\ldots a_{11}}G_{[a_1, a_2\ldots a_{11}], b_1b_2}$
transforms in the same way as  the usual symmetrised gravity field $h_{(ab)}$. The  analogue of the field $h_{b_1b_2}$ being the 
Cartan form $\ \ \ \ \ \ \ \quad$ $\epsilon ^{a_1 a_2\ldots a_{11}} G_{[a_1, a_2\ldots a_{11}], b_1b_2}$  and the transformations parameter $\xi_b$ being the parameter $\quad$ $\epsilon ^{a_1 a_2\ldots a_{11}}\partial_{a_1} \Lambda _{a_1 a_2\ldots a_{11}, b }$. In the considerations that follow the eleven indices will play no role and go along for the ride and so we will simply ignore them in the discuss. 
\par
Examining equation (4.24) we observe that it contains the object  $E^{(2)}{}_{a_1  a_2\ldots a_{11} ,   b c, d}  \equiv -2\partial _{[b|} G_{[a_1 , a_2\ldots a_{11}] ,  |c]d } $ and under the above gauge transformation this transforms as 
$$
\delta\, E^{(2)}{}_{c_1...c_{11},\,b_1b_2\,a} = -\,\partial_a\,\partial_{[b_1}\,\partial_{[c_1}\,\Lambda_{c_2...c_{11}],\,b_2]}\equiv 
\partial_a \hat \Lambda_{[c_1 c_2...c_{11}],\,b_1 b_2}.
\eqno(6.5)$$
We observe that $E^{(2)}{}_{c_1...c_{11},\,b_1b_2,a}$ changes in the same way as the usual  spin connection  but by  a composite Lorentz transformation with parameter $\hat \Lambda_{[c_1 c_2...c_{11}],\,b_1 b_2}$. In fact $E^{(2)}{}_{c_1...c_{11},\,b_1b_2\,a,}$ contains  the analogue of the first two terms of the spin connection as given in equation (3.7). 
\par
The simplest  gauge invariant object is 
$$
 E^{(3)}{}_{c_1...c_{11},\,a_1a_2,\,b_1b_2} = -{1\over 2}(\partial_{a_1} E^{(2)}{}_{c_1...c_{11},\,b_1b_2,\,a_2} - \partial_{a_2}\, E^{(2)}{}_{c_1...c_{11},\,b_1b_2,a_1})
$$\
$$
= 2\,\partial_{[a_1}\,\partial_{[b_1}\,G_{[c_1,\,c_2...c_{11}], a_2]b_2]},
\eqno(6.6)$$
\par
We identify $  E^{(3)}{}_{c_1...c_{11},\,a_1a_2,\,b_1b_2}$ as the analogue of the Riemann tensor. Indeed starting from the observation below equation (6.4) we could have constructed the analogue of the Christoffell symbol and then the analogue of the Riemann tensor however the result is the same as the object $ E^{(3)}{}_{c_1...c_{11},\,a_1a_2,\,b_1b_2}$ of equation (6.6). 
\par
Thus  the above  gauge invariance considerations suggest that  the field $A_{a_1\ldots a_{10}, b_1b_2}$ should obey the equation of motion 
$$
 E^{(3)}{}_{c_1...c_{11},\,a_1a_2,\,b_1b_2}= 2\,\partial_{[a_1}\,\partial_{[b_1}\,G_{[c_1,\,c_2...c_{11}],\,a_2]b_2]}
=0 
\eqno(6.7)$$
\par
In order to confirm this field equation we now consider its $E_{11}$ transformations and show that it  leads to the other equations that we have also found without having to require any modulo transformations.  The $I_c(E_{11})$ transformations of the field $A_{a_1\ldots a_{10}, b_1b_2}$  into the fields of level three can be found from equation (2.9) using the commutators of appendix A. One finds that 
$$
\delta\,G_{a_1...a_{10},\,b_1b_2} = {24\over 11}\,G_{[a_1...a_7\left(b_1,\,b_2\right)}\,\Lambda_{a_8a_9a_{10}]} - {24\over 11}\,G_{[a_1...a_8,\,(b_1}\,\Lambda_{b_2)a_9a_{10}]}+\ldots ,
\eqno(6.8)$$
where $+\ldots$ indicates the presence of level five terms which we will not consider. Under this transformation the Cartan form of interest transforms as 
$$
\delta\,G_{[a_1,\,a_2...a_{11}],\,b_1b_2} = {27\over 11}\,G_{[[a_1,\,a_2...a_8\left(b_1],\,b_2\right)}\,\Lambda_{a_9a_{10}a_{11}]}
$$
$$
-\,{24\over 11}\,G_{[a_1,\,a_2...a_9,\,(b_1}\,\Lambda_{b_2)a_{10}a_{11}]} - {3\over 11}\,G_{\left(b_1,\,[a_1...a_8,\,|b_2\right)|}\,\Lambda_{a_9a_{10}a_{11}]}.
\eqno(6.9)$$
A calculation shows that the Riemann tensor transforms as 

$$
\delta\, E^{(3)}{}_{c_1...c_{11},\,a_1a_2,\,b_1b_2} = -\,{15\over 11}\,\partial_{[a_1}\,\partial_{[b_1}\,\left(G_{[c_1,\,c_2...c_9|,\,b_2]}\,\Lambda_{|c_{10}c_{11}]a_2]} + G_{[c_1,\,c_2...c_9|,\,a_2]}\,\Lambda_{|c_{10}c_{11}]b_2]}\right)
$$
$$
= -\,{60\over 11\cdot 11!}\,\varepsilon_{c_1...c_{11}}\,\partial_{[a_1}\,\partial_{[b_1}\,\left( E^{(1)}{}_{a_2],\,d_1d_2}\,\Lambda_{b_2]\,d_1d_2} +  E^{(1)}{}_{b_2],\,d_1d_2}\,\Lambda_{a_2]\,d_1d_2}\right)
$$
$$
+\,{60\over 11\cdot 11!}\,\varepsilon_{c_1...c_{11}}\,\partial_{[a_1}\,\partial_{[b_1}\,\left(\omega_{a_2],\,d_1d_2}\,\Lambda_{b_2]\,d_1d_2} +  \omega_{b_2]\,d_1d_2}\,\Lambda_{a_2],\,d_1d_2}\right).
\eqno(6.10)$$
To go from the first line of this equation to the second line we have used the gravity-dual gravity relation of equation (4.9). As a result the first two terms of equation (6.10) contain the gravity-dual gravity relation and the  $\omega_{\,a,\,b_1b_2}$ that appears  is the usual spin connection of equation (3.8). We note that although this last relation only holds modulo local Lorentz transformations it is straightforward to show that the duality relation that occurs in this equation in just such a way that  these transformations are eliminated. The last two terms of equation (6.10) can be cancelled by adding terms to $ E^{(3)}{}_{c_1...c_{11},\,a_1a_2,\,b_1b_2}$ which contain derivatives with respect to the higher level coordinates. This can be seen by reading the discussion at the end of section two and writing  equation  (6.10) in the form 
$$
\delta\, E^{(3)}{}_{c_1...c_{11},\,a_1a_2,\,b_1b_2} = -\,{15\over 11}\,\partial_{[a_1}\,\partial_{[b_1}\,\left(G_{[c_1,\,c_2...c_9|,\,b_2]}\,\Lambda_{|c_{10}c_{11}]a_2]} + G_{[c_1,\,c_2...c_9|,\,a_2]}\,\Lambda_{|c_{10}c_{11}]b_2]}\right)
$$
$$
= -\,{60\over 11\cdot 11!}\,\varepsilon_{c_1...c_{11}}\,\partial_{[a_1}\,\partial_{[b_1}\,\left( E^{(1)}{}_{a_2],\,}{}^{d_1d_2}\,\Lambda_{b_2]\,d_1d_2} +  E^{(1)}{}_{b_2]\,}{}^{d_1d_2}\,\Lambda_{a_2],\,d_1d_2}\right)
$$
$$
-\,{120\over 11\cdot 11!}\,\varepsilon_{c_1...c_{11}}\,\partial_{[a_1}\,\partial_{[b_1}\,\left(\partial_{d_1}\,h_{\left(d_2|a_2]\right)}\,\Lambda_{b_2]\,}{}^{d_1d_2} + \partial_{d_1}\,h_{\left(d_2|b_2]\right)}\,\Lambda_{a_2]\,}{}^{d_1d_2}\right).
\eqno(6.11)$$
  Thus we have found that the equation of motion of equation (6.7) does indeed vary precisely into the previous equations  and hence we have found an equation, that does not involve any modulo transformations, and  whose $E_{11}$ variation gives our other equations without the use of modulo transformations.  As we have explained above this equation is also  gauge invariant. As a result,  equation (6.7) is the field equation for the field $A_{a_1\ldots a_{10}, b_1b_2}$. This also implies that equation (4.24) only holds modulo certain transformations and these can be found following the discussion at the end of section five. 
\par
From equation (6.10) one can also find the $E_{11}$  variation of the analogue of the Ricci tensor $R_{c_1\ldots c_{11}, ab}$ by taking a contraction. It follows that this will also vary into the previous field equations. In principle one could take our equation to be that the analogue of the Ricci tensor to vanish rather than the Riemann tensor is zero. However, this is unlikely to be the case in view of the fact the vanishing of the Riemann tensor satisfies all the above requirements. 
 \par
We now consider the consequences of the field equation (6.7) for the field 
$A_{a_1\ldots a_{10}, b_1b_2}$ which essentially states that the analogue of the Riemann tensor vanishes. We recall that if the Riemann tensor vanishes then one can find a coordinate system in which the spacetime is flat. Applying this to our setting  we can conclude that there exists a gauge in which 
 $$
G_{[a_1 , a_2\ldots a_{11}] ,  bc} = m
\epsilon _{a_1  a_2\ldots a_{11} }\eta_{bc }
\eqno(6.12)$$
where $m$ is a parameter. This makes it clear that the field $A_{a_1  a_2\ldots a_{10}, bc }$ carries no degrees of freedom. Thus even though we have a field equation with three derivatives $E_{11}$ has found a way to ensure that there are no additional  degrees of freedom from this level four field. Since the full equations will be general coordinate invariant the right hand side of equation (6.12) will be promoted to the same expression, provided if we use tangent indices,  plus terms bilinear and higher  in the Cartan forms of other fields. 
\par
We now  consider the dimensional reduction to ten dimensions, that is, to the IIA theory. We find the eleven-dimensional  field $A_{\hat a_1\ldots \hat a_{10}, \hat b_1 \hat b_2}$, where now $\hat a, \hat b,\ldots =0,1\ldots , 10$, gives rise to the following fields  $A_{a_1\ldots a_{10}, b_1b_2}$   $A_{a_1\ldots a_{9 }, b_1 b_2}$,  $A_{a_1\ldots a_{10}}$ and 
$A_{a_1\ldots a_{9}}$ in ten dimensions. In listing these fields we have taken into account the irreducibility condition $A_{[\hat a_1\ldots \hat a_{10}, \hat b_1] \hat b_2}=0$ valid before dimensional reduction. We see from equation (4.24) that the  field $A_{a_1\ldots a_{9}}$ obeys the equation 
$$ F_{a_1\ldots a_{10}}\propto m\epsilon _{a_1\ldots a_{10}}
\eqno(6.13)$$
where $F_{a_1\ldots a_{10}}= \partial_{[a_1}A_{a_2\ldots a_{10}]}$.  
 If were to assume that this field occurs in the action in the usual way this would lead to a theory with a cosmological constant which must be Romans theory. From the viewpoint of the non-linear realisation one should find that the square of this field strength appears in the energy momentum tensor on the right-hand side of the Einstein equation. It would be interesting to show this explicitly.


\medskip 
{\bf 7. Discussion}
\medskip 
In this paper we have found, at the linearised level, the  equations of motion that result from the $E_{11}\otimes_s l_1$ non-linear realisation up to and including level four. A general pattern can be discerned. One finds a set of $E_{11}$ invariant equations that are first order in derivatives. These equations are often but not always duality equations in that they contain two different fields; they also only hold modulo certain transformations, the one exception being the lowest level such equation.   From these equations  one can deduce, by taking derivatives, a set of equations that are second order in derivatives. These equations with two derivatives rotate into themselves and the equations with one derivative under $E_{11}$ transformations. At lowest levels, these equations, which have  two derivatives,  hold exactly that is without being modulo certain transformations and they contain only a single field. They are the    familiar  second order equations that  account  for the degrees of freedom of the eleven dimensional supergravity theory. However, at level four one finds two equations that are second order in derivatives; one of these holds exactly and relates one of the level four fields  to the three form field and so is a novel kind of duality equation, while the other only contains a single level four field but it only holds modulo certain transformations. By taking derivatives of this latter equation one finds an equations that is third order in derivatives that holds exactly and rotates under $E_{11}$ transformations  into the other equations. This later equation, when reduced to ten dimensions, leads to Roman's theory and so it provides an eleven dimensional origin for this theory.  The web of $E_{11}$ invariant  equations  are given in the table one in section one. We note that even though we have calculated up to level four in the fields we have not required any section condition to verify the $E_{11}$ invariance and we see no need for such a condition. 
\par
Thus we find that as one goes to higher and higher levels one finds that one needs more and more derivatives to find an equation that holds exactly. These exact  equations can be derived from the equations with less derivatives, including equations with only one derivative. The set of all equations rotate into themselves under $E_{11}$ transformations which is in fact how they were derived. The pattern is that the exact equation for a given field possess one derivative for each block of antisymmetric indices that the field possess. Thus if a field has $n$ blocks of indices the exact equation will contain $n$ derivatives and this equation can be derived from equations with lower number of derivatives but these equations only hold modulo certain transformations. This rule applies to all the equations derived in this paper and in particular it applies to the very lowest level, and very familiar, duality equation which expresses the duality between the three form and six form fields which  holds  exactly and is gauge invariant. We are used to think of the dynamics as being given by exact equations but $E_{11}$ produces instead a web of equations only the ones with the most derivatives being exact. In this sense the dynamics that the non-linear realisation leads to is unconventional. 
\par
As noted in the introduction the $E_{11}\otimes_s l_1$ non-linear realisation contains an infinite number of fields which are dual to the fields which we usually use to describe the degrees of freedom of supergravity, see equation (1.1). For the three,  six form  and their higher level dual fields a system of equations of the type found in this paper has been found on general grounds. That is a series of  duality equations that are first order in derivatives and only hold modulo certain transformations and from these one can derive exact equations by taking derivatives [11]. In this paper we have found that the first few such equations are contained in   the web of  $E_{11}$ invariant equations and as a result they really do transform into each other under duality transformations as these are part of the $E_{11}$ symmetry. 
\par
Although we have not assumed that the equations are gauge invariant the exact equations that we have derived are gauge invariant. Thus it  seems that gauge invariance is a consequence of $E_{11}$ invariance, at least up to the level investigated. It follows that the geometry that underlies the $E_{11}\otimes_s l_1$ non-linear realisation is very far from being a Riemannian geometry. It would be   of interest to understand how to construct the equation that result from the non-linear realisation more systemically as the complexity increases greatly as one goes to higher levels. It could be that one can find the geometry hidden in the non-linear realisation or use the unfolding techniques that occur in higher spin theories whose equations have several similarities to the equations we find in this paper. 
\par
We note that the level four field which possess a block of eleven antisymmetrised indices did not occur in the equations of motion and one could speculate that this could be a general feature, namely that  fields with blocks of eleven antisymmetrised indices play no role in the dynamics.  
\par
The equations of motion involving the lowest level fields given in this paper have previously been found at the full non-linear level. These include the equation for the three form and the metric; they are precisely the equations of eleven dimensional supergravity [7,8]. However, it would be good to find the non-linear analogues of the equation of motion  (3.11) for the dual graviton and the first order duality relation between the graviton and the dual graviton, equation (4.9). This is currently being studied. 

\par
In carrying out this calculation we began from the $E_{11}$ Dynkin diagram and its vector representation and calculated the consequences of the non-linear realisation $E_{11}\otimes_s l_1 $. The only other assumption was that the local subgroup in the non-linear realisation was the Cartan involution invariant subgroup $I_c(E_{11})$. The resulting theory has all its fields and coordinates specified, the former  just correspond to the Borel subalgebra generators of $E_{11}$ and the later to the elements of the vector representation. The dynamics is just  the set of $E_{11}$ invariant equations with the fewest number of derivatives. Thus the bosonic sectors of the maximal supergravity theories  follow from this construction, at low levels,  in a unique way and one can even say that they are encoded in the $E_{11}$ Dynkin diagram. Indeed by truncation of the maximal supergravity theories one can find all the bosonic sectors of the other non-maximal supergravity theories and so essentially much  of supergravity  can be derived from the $E_{11}$ Dynkin diagram. The equations of motion that emerge correctly describe not only the dynamics of the usual fields but also that of the  higher level fields where this has been tested, such as in this paper. This requires very many conspiracies and it could have failed at any point. We encourage the reader to try some of the calculations in this and the papers [7,8]  and see this for him or herself. This, and most previous papers,   concern only the bosonic fields of supergravity and while one can include fermions [19] along the lines of that first given in the $E_{10}$ context [20] it would be interesting to derive these results from a deeper perspective. 

\medskip
{\bf Appendix A}
\medskip

The $E_{11}$ algebra in eleven dimensions has been found up to level three in previous papers on $E_{11}$ and the result is given in the book of reference  [17]. In this paper we are using some of the commutators that involve level four  generators in both the $E_{11}$ algebra  and the $l_1$ part of the semi-direct product algebra $E_{11}\otimes _S l_1$. The full results will be published elsewhere [21] but here we  present the ones we needed in this paper  in this appendix. These commutators were found in collaborations with  Nikolay Gromov using  Wolfram Mathematica. The level $4$ generators appear in the commutator of level three generators  with the level one generator and these commutators are given by [22]
$$
\left[R^{a_1a_2a_3},\,R^{b_1...b_8,\,c}\right] = {3\over 2}\,R^{b_1...b_8[a_1,\,a_2a_3]c} - {1\over 6}\,R^{b_1...b_8c,\,a_1a_2a_3}
$$
$$
+\,R^{b_1...b_8[a_1a_2,a_3]c} + R^{a_1a_2a_3b_1...b_8,\,c} - {1\over 3}\,R^{b_1...b_8c[a_1a_2,\,a_3]},
\eqno(A.1)$$
The commutators of level four generators with the level $-1$ generator are given by 
$$
\left[ R_{a_1a_2a_3},\,R^{b_1...b_9,\,c_1c_2c_3} \right] = 189\,\delta_{\,a_1a_2\,a_3}^{[c_1c_2[b_1}\,R^{b_2...b_9],\,c_3]} + 432\,\delta_{\,a_1\,a_2a_3}^{[c_1[b_1b_2}\,R^{b_3...b_9]c_2,\,c_3]} 
$$
$$
+ 252\,\delta_{\,a_1a_2a_3}^{[b_1b_2b_3}\,R^{b_4...b_9][c_1c_2,\,c_3]},
\eqno(A.2)$$
$$
\left[ R_{a_1a_2a_3},\,R^{b_1...b_{10},\,c_1c_2} \right] = {405\over 2}\,\delta_{\,a_1a_2a_3}^{[b_1b_2(c_1}\,R^{b_3...b_{10}],\,c_2)} - 180\,\delta_{\,a_1a_2a_3}^{[b_1b_2b_3}\,R^{b_4...b_{10}](c_1,\,c_2)},
\eqno(A.3)$$
$$
\left[ R_{a_1a_2a_3},\,R^{b_1...b_{11},\,c} \right] = \,{495\over 4}\,\delta_{\,a_1a_2a_3}^{[b_1b_2b_3}\,R^{b_4...b_{11}],\,c}.
\eqno(A.4)$$
\par
In order to determine in section six the gauge transformation of the $A_{a_1\ldots a_{10}, b_1b_2}$ field  we require certain  of the commutators of the $E_{11}$ generators with those level four generators in  the $l_1$ representation. There are six different $l_1$ generators at level four:
$$
Z^{a_1...a_8,\,b_1b_2b_3},\quad Z^{a_1...a_{11}},\quad Z^{a_1...a_9,\,b_1b_2},\quad \hat  Z^{a_1...a_9,\,b_1b_2},\quad Z^{a_1...a_{10},\,b}_{\left(1\right)},\quad Z^{a_1...a_{10},\,b}_{\left(2\right)},
\eqno(A.5)$$
where indexes $\left(1\right)$ and $\left(2\right)$ indicate that generator $Z^{a_1...a_{10},\,b}$ has multiplicity $2$ and these generator satisfies the condition $\hat  Z^{a_1...a_9,\,b_1b_2}_{(n)}= \hat  Z^{a_1...a_9,\, (b_1b_2 )}_{(n)}$ for $n=1,2$. However, we will require only the  last three generators and it will also turn out that we also only require those that have the generators $P_a$ on the right-hand side. The corresponding commutators are given by 
$$
\left[R_{a_1...a_{10},\,c_1c_2},\,\hat Z^{b_1...b_9,\,d_1d_2}\right] = {189\cdot 12!\over 88}\,\left(\delta_{\left(c_1c_2\right)}^{\,d_1d_2}\,\delta_{[a_1...a_9}^{\,b_1...b_9}\,P_{a_{10}]} + 2\,\delta_{\left(c_1\,\,c_2\right)}^{[b_1(d_1}\,\delta_{[a_1\,\,.\,\,.\,\,.\,\,a_9}^{\,d_2)b_2...b_9]}\,P_{a_{10}]}\right),
\eqno(A.6)$$
$$
\left[R_{a_1...a_{10},\,c_1c_2},\,Z^{b_1...b_{10},\,d}_{\left(1\right)}\right] = {105\cdot 10!\over 44}\,(\delta_{a_1...a_{10}}^{b_1...b_{10}}\,\delta_{(c_1}^{\,d}\,P_{c_2)} 
$$
$$+ \delta_{\,\,a_1\,\,.\,.\,.\,\,a_{10}}^{[b_1...b_9|d|}\,\delta_{(c_1}^{\,b_{10}]}\,P_{c_2)} + \delta^{\,b_1\,\,.\,\,.\,\,.\,\,b_{10}}_{[a_1...a_9(c_1}\,\delta_{c_2)}^{\,d}\,P_{a_{10}]}),
\eqno(A.7)$$
$$
\left[R_{a_1...a_{10},\,c_1c_2},\,Z^{b_1...b_{10},\,d}_{\left(2\right)}\right] = {5\cdot 10!\over 4}\,(\delta_{a_1...a_{10}}^{b_1...b_{10}}\,\delta_{(c_1}^{\,d}\,P_{c_2)} 
$$
$$+ \delta_{\,\,a_1\,\,.\,.\,.\,\,a_{10}}^{[b_1...b_9|d|}\,\delta_{(c_1}^{\,b_{10}]}\,P_{c_2)} + \delta^{\,b_1\,\,.\,\,.\,\,.\,\,b_{10}}_{[a_1...a_9(c_1}\,\delta_{c_2)}^{\,d}\,P_{a_{10}]}).
\eqno(A.8)$$

\medskip
{\bf Appendix B}
\medskip
Now we show that the linearised  dual graviton equation, found in this paper, that is, equation (3.10),  $E^{(2)}{}_{a_1...a_8,\,b}=0$,  forms an irreducible representation of  SL(11). We recall that  
$$
E^{(2)}{}_{a_1...a_8,\,b} = -\,{1\over 4}\,\partial_{[c}\,G_{[c,\,a_1...a_8],\,b]} = -\,{1\over 4}\,\partial_{[c}\,\partial_{[c}\,A_{a_1...a_8],\,b]}
\eqno(B.1)$$
Expanding out the anti-symmetries we can write this object as 
$$
E^{(2)}{}_{a_1...a_8,\,b} = \,-\,{1\over 72}\,\partial^2\,A_{a_1...a_8,\,b} + {1\over 9}\,\partial_b\,\partial_{[a_1}\,A_{a_2...a_8]c,\,}{}^{c} 
$$
$$+ {1\over 72}\,\partial^c\,\partial_b\,A_{a_1...a_8,\,c} - {1\over 9}\,\partial^c\,\partial_{[a_1}\,A_{a_2...a_8]c,\,b}
\eqno(B.2)$$
and as a result the completely antisymmetrised object satisfies the equation 
$$
E^{(2)}{}_{[a_1...a_8,\,b]} = \,-\,{1\over 72}\,\partial^2\,A_{[a_1...a_8,\,b]} + {1\over 9}\,\partial_{[b}\,\partial_{a_1}\,A_{a_2...a_8]c,\,c} 
$$
$$- {5\over 36}\,\partial_c\,\partial_{[c}\,A_{a_1...a_8,\,b]} - {1\over 72}\,\partial^2\,A_{[a_1...a_8,\,b]} = 0.
\eqno(B.3)$$
In deriving this result we have used the following identity 
$$
\partial_{[b}\,A_{a_1...a_8],\,c} - 8\,\partial_{[a_1}\,A_{a_2...a_8|c|,\,b]} = 10\,\partial_{[b}\,A_{a_1...a_8,\,c]} + \partial_{c}\,A_{[a_1...a_8,\,b]}.
\eqno(B.4)$$
on the last two terms in equation (B.2). Examining equation (B.3) we find that the linearised dual graviton equation of motion does belong to an irreducible representation of SL(11) as it satisfies the condition 
$$
E^{(2)}{}_{[a_1...a_8,\,b]} = 0
\eqno(B.5)$$
We have used   the irreducibility property of the $A_{a_1...a_8,\,b}$ field ($A_{[a_1...a_8,\,b]} = 0$) and the symmetry of the derivatives. 
This is to be expected as the field equation for a given field often belongs to the same irreducible representation as the field since it is usually derivable from an action that is bilinear in the field. We also note that  the field that appears in $E_{11}$ is just the irreducible part. The considerations of this, and previous papers,  dispel the doubts that only this part would not be enough to describe  gravity using this field.  


\medskip

\medskip
{\bf Note added}
\medskip
In this note added we will fill in some of the gaps concerning the equations satisfied by the field $A_{a_1\ldots a_{10}, b_1b_2}$. In particular  we will find the $I_c(E_{11})$ variations of the equations involving this field and show that they give the other equations plus terms involving modulo contributions. In particular, we find the modulo terms  for the equations which are first and second order in derivatives while   the equation that is  third order in derivatives varies exactly into the other equations. In this way we will derive precisely what are the modulo terms up to which  the equation involving $A_{a_1\ldots a_{10}, b_1b_2}$ hold. 
\par
In section five, by varying the gravity-dual gravity equation,  we derived  equation (5.9) which was first order in derivatives and 
involved the field $A_{a_1\ldots a_{10}, b_1b_2}$. We find that the $I_c(E_{11})$ variation of this equation is given by 
$$
\delta {\cal E}_{c_1...c_{11},\,a_1a_2} ^{(1)}= -\,{60\over 11\cdot 11!}\,\varepsilon_{c_1...c_{11}}\,E^{(1)}_{(a_1|,\,d_1d_2}\,\Lambda_{|a_2)}{}^{d_1d_2}  - \varepsilon_{c_1...c_{11}}\,\partial_{(a_1}\,\tilde{\Lambda}_{a_2)},
\eqno(N.1)$$
where $E^{(1)}_{a,\,b_1b_2}$ is the gravity-dual gravity relation of equation (5.2), or equation (4.7),  
$$
{\cal E}_{c_1...c_{11},\,a_1a_2} ^{(1)}\equiv G_{[c_1,\,c_2...c_{11}],\,a_1a_2} + {20\over 11\cdot 11!}\,\varepsilon_{c_1...c_{11}}\,G^d{}_{\left(a_1,\,(a_2\right)d)},
\eqno(N.2)$$
and $\partial_{(a_1}\,\tilde{\Lambda}_{a_2)}$ is the modulo term given by
$$
\partial_{(a_1}\tilde{\Lambda}_{a_2)} = -\,{60\over 11\cdot 11!}\,\left(G_{(a_1,\,|d_1d_2|}\,\Lambda_{a_2)}{}^{d_1d_2} + {1\over 20}\,\varepsilon^{d_1...d_{11}}\,G_{(a_1,\,|d_1...d_8|,\,a_2)}\,\Lambda_{d_9d_{10}d_{11}}\right).
\eqno(N.3)$$
One can extract off the derivative from both sides of this equation to find a much more complicated expression in terms of the $E_{11}$ fields. We note that $\Lambda_{a}{}^{d_1d_2}$ is a constant in this equation as we are working at the linearised level. As we did when varying the other equations of motion in this paper we have added  terms which  involve  the derivatives of the level one coordinates to the object being varied. 
\par
Looking at equation (N.2) we conclude that equation (5.9) holds in the following sense 
$$
{\cal E}_{c_1...c_{11},\,a_1a_2} ^{(1)}\dot =0 , \quad {\rm meaning }\quad 
{\cal E}_{c_1...c_{11},\,a_1a_2} ^{(1)}- \varepsilon_{c_1...c_{11}}\,\partial_{(a_1}\,\hat {\Lambda}_{a_2)}=0
\eqno(N.4)$$
for arbitrary $\hat {\Lambda}_{a_2}$. 
This analysis justifies and makes precise the meaning of equation (5.9)
\par
 In section four we found equation (4.24)  that was second order in derivatives and was derived from the variation given in  equation (4.21). The $I_c(E_{11})$  variation of the object in equation (4.24) is given by 
$$
\delta {\cal E}^{(2)}_{c_1...c_{11},\,a,\,b_1b_2} = {60\over 11\cdot 11!}\,\varepsilon_{c_1...c_{11}}\,\partial_{[b_1|}\,\left(E^{(1)}_{|b_2],\,d_1d_2}\,\Lambda_a{}^{d_1d_2} + E^{(1)}_{a,\,d_1d_2}\,\Lambda_{|b_2]}{}^{d_1d_2}\right)
$$
$$
+ \varepsilon_{c_1...c_{11}}\,\partial_a\,\partial_{[b_1}\,\tilde{\Lambda}_{b_2]},
\eqno(N.5)$$
where 
$$
{\cal E}^{(2)}_{c_1...c_{11},\,a,\,b_1b_2} = -\,2\,\partial_{[b_1}\,G_{[c_1,\,c_2...c_{11}],\,b_2]a} - {20\over 11\cdot 11!}\,\varepsilon_{c_1...c_{11}}\,\left(\partial^d{}_a\,G_{[b_1,\,\left(b_2]d\right)} - \partial^d{}_{[b_1}\,G_{b_2],\,\left(ad\right)}\right).
\eqno(N.6)$$
and  $\partial_a\,\partial_{[b_1}\,\tilde{\Lambda}_{b_2]}$ is the modulo term given by
$$
\partial_{[b_1}\tilde{\Lambda}_{b_2]} = -\,{60\over 11\cdot 11!}\,\left(G_{[b_1,\,|d_1d_2|}\,\Lambda_{b_2]}{}^{d_1d_2} + {1\over 20}\,\varepsilon^{d_1...d_{11}}\,G_{[b_1,\,|d_1...d_8|,\,b_2]}\,\Lambda_{d_9d_{10}d_{11}}\right).
\eqno(N.7)$$
We observe that equation (4.24)   varies into the other equations but also has a modulo term. Indeed examining the above equation we conclude that equation (4.24) holds in the sense  
$$
{\cal E}^{(2)}_{c_1...c_{11},\,a,\,b_1b_2}\dot =0 ,\quad {\rm meaning }\quad {\cal E}^{(2)}_{c_1...c_{11},\,a,\,b_1b_2}+ \varepsilon_{c_1...c_{11}}\,\partial_a\,\partial_{[b_1}\,\hat{\Lambda}_{b_2]}=0
\eqno(N.8)$$
for arbitrary $\hat{\Lambda}_{b}$. 
\par
We note that the objects in equations (N.2) and (N.5) are related by a projector 
$$
E^{(2)}_{a_1...a_{11},\,b,\,c_1c_2} \equiv  \left(\sigma\,E^{(1)}\right)_{a_1...a_{11},\,b,\,c_1c_2} = -\,2\,\partial_{[c_1|}\,E^{(1)}_{a_1...a_{11},\,|c_2]b} = -\,2\,\partial_{[c_1}\,G_{[a_1,\,a_2...a_{11}],\,c_2]b}
\eqno(N.9)$$
where we neglect the terms with derivatives with respect to the level one coordinates. 
\par
We can act with  a projector,  which is first order in derivatives, on the 
 quantity of  equation (N.8) to  find a relation whose variation holds exactly. 
We consider the object 
$$
E^{(3)}_{a_1...a_{11},\,b_1b_2,\,c_1c_2} \equiv  \left(\rho\,E^{(2)}\right)_{a_1...a_{11},\,b_1b_2,\,c_1c_2} = -\,\partial_{[b_1}\,E^{(2)}_{|a_1...a_{11}|,\,b_2],\,c_1c_2}
$$
$$
= 2\,\partial_{[b_1}\,\partial_{[c_1}\,E^{(1)}_{|a_1...a_{11}|,\,c_2]b_2]} = 2\,\partial_{[b_1}\,\partial_{[c_1}\,E^{(1)}_{[a_1,\,a_2...a_{11}],\,c_2]b_2]}.
\eqno(N.10)$$
As the equation shows it can also be found by acting with  a projection with two derivatives on the object we considered at the beginning of this section. 
The $I_c(E_{11}) $variation is given by 
$$
\delta {\cal E}^{(3)}_{c_1...c_{11},\,a_1a_2,\,b_1b_2} = -\,{60\over 11\cdot 11!}\,\varepsilon_{c_1...c_{11}}\,\partial_{[a_1|}\,\partial_{[b_1|}\,\left( E^{(1)}_{|a_2],\,d_1d_2}\,\Lambda_{|b_2]}{}^{d_1d_2} + E^{(1)}_{|b_2],\,d_1d_2}\,\Lambda_{|a_2]}{}^{d_1d_2}\right),
\eqno(N.11)$$
where
$$
{\cal E}^{(3)}_{c_1...c_{11},\,a_1a_2,\,b_1b_2} = 2\,\partial_{[a_1}\,\partial_{[b_1}\,G_{[c_1,\,c_2...c_{11}],\,b_2]a_2]}
$$
$$
- {20\over 11\cdot 11!}\,\varepsilon_{c_1...c_{11}}\,\left(\partial^d{}_{[b_1}\,\partial_{b_2]}\,E^{(1)}_{[a_1,\,\left(a_2]d\right)} + \partial^d{}_{[a_1}\,\partial_{a_2]}\,E^{(1)}_{[b_1,\,\left(b_2]d\right)}\right).
\eqno(N.12)$$
We see that ${\cal E}^{(3)}_{c_1...c_{11},\,a_1a_2,\,b_1b_2} $ does indeed vary into the other equations, in particular the gravity-dual gravity relation, without any modulo terms and so we can take it to hold exactly, that is,  
$$
{\cal E}^{(3)}_{c_1...c_{11},\,a_1a_2,\,b_1b_2} =0
\eqno(N.13)$$
We observe that the modulo transformations of the gravity-dual gravity relation are eliminated by the way this object appears in ${\cal E}^{(3)}_{c_1...c_{11},\,a_1a_2,\,b_1b_2} $.
The result of equation (N.13) is the same as that of equation (6.7). However, we arrived at this last equation by a different route, that is, by finding an equation that was  the gauge transformation rather than using the $I_c(E_{11})$. Hence we confirm  that the field $A_{a_1\ldots a_{10}, b_1b_2}$ obeys an exact equation that is third order in derivatives. We recall that we found in section six that this equation lead to no degrees of freedom. 
\par
It was shown in section five that the modulo terms in the equations of motion are closely related to gauge transformations.  Examining equations (N.4) and (N.8) and the gauge transformations of equations (6.4) and (6.5) we find that the same conclusion holds for the equations of motion of the field $A_{a_1\ldots a_{10}, b_1b_2}$.

\medskip

\medskip
{\bf {Acknowledgements}}
\medskip
We wish to thank Nikolay Gromov  and Nicolas Boulanger for their help. We also wish to thank the STFC for support from Consolidated grant number ST/J002798/1 and Alexander Tumanov wishes to thanks King's College  for the support provided by his  Graduate School International Research Studentship. 
\medskip
{\bf {References}}
\medskip
\item{[1]} P. West, {\it $E_{11}$ and M Theory}, Class. Quant.  
Grav.  {\bf 18}, (2001) 4443, hep-th/ 0104081. 
\item{[2]} P. West, {\it $E_{11}$, SL(32) and Central Charges},
Phys. Lett. {\bf B 575} (2003) 333-342,  hep-th/0307098. 
\item{[3]}I. Schnakenburg and  P. West, {\it Kac-Moody   
symmetries of
IIB supergravity}, Phys. Lett. {\bf B517} (2001) 421, hep-th/0107181.
\item{[4]} P. West, {\it The IIA, IIB and eleven dimensional theories 
and their common
$E_{11}$ origin}, Nucl. Phys. B693 (2004) 76-102, hep-th/0402140. 
\item{[5]}  F. ÊRiccioni and P. West, {\it
The $E_{11}$ origin of all maximal supergravities}, ÊJHEP {\bf 0707}
(2007) 063; ÊarXiv:0705.0752.
\item{[6]} ÊF. Riccioni and P. West, {\it E(11)-extended space time
and gauged supergravities},
JHEP {\bf 0802} (2008) 039, ÊarXiv:0712.1795.
\item{[7]} A. Tumanov and P. West, {\it E11 must be a symmetry of strings and branes },  Phys. Lett. {\bf  B759 }Ê(2016),  663, arXiv:1512.01644. 
\item{[8]} A. Tumanov and P. West, {\it E11 in 11D}, Phys.Lett. B758 (2016) 278, arXiv:1601.03974. 
\item{[9]} P. West, {\it Very Extended $E_8$ and $A_8$ at low
levels, Gravity and Supergravity}, Class.Quant.Grav. {\bf 20} (2003)
2393, hep-th/0212291.
\item{[10]} ÊF. Riccioni and P. West, {\it Dual fields and $E_{11}$},   Phys.Lett.B645 (2007) 286-292,  hep-th/0612001; F. Riccioni, D. Steele and P. West, {\it Duality Symmetries and $G^{+++}$ Theories},  Class.Quant.Grav.25:045012,2008,  arXiv:0706.3659. 
\item{[11]}  N. Boulanger,  P, Sundell and P. West, {\it Gauge fields and infinite chains of dualities},  , JHEP 1509 (2015) 192,  arXiv:1502.07909. 
\item{[12]} L. J. Romans, {\sl Massive $N=2A$ supergravity in ten
  dimensions}, Phys. Lett. {\bf B 169} (1986) 374.
\item{[13]}  A.~Kleinschmidt, I.~Schnakenburg and P.~West, {\it Very-extended Kac-Moody algebras and their interpretation at low  levels}, Class.\ Quant.\ Grav.\  {\bf 21} (2004) 2493 [arXiv:hep-th/0309198].; P.~West, {\it E(11), ten forms and supergravity},  JHEP {\bf 0603} (2006) 072,  [arXiv:hep-th/0511153]. 
\item {[14]} P. West, {\it Generalised Geometry, eleven dimensions
and $E_{11}$}, JHEP 1202 (2012) 018, arXiv:1111.1642.  
\item {[15]} P. West, {\it A brief review of E theory}, Proceedings of Abdus Salam's 90th  Birthday meeting, 25-28 January 2016, NTU, Singapore, Editors L. Brink, M. Duff and K. Phua, World Scientific Publishing and IJMPA, {\bf Vol 31}, No 26 (2016) 1630043,  arXiv:1609.06863. 
\item{[16]} P. West, {\it Dual gravity and E11  },  arXiv:1411.0920. 
\item{[17]} P. West, {\it Introduction to Strings and Branes}. Cambridge University Press, 2012. 
\item{[18]} P. West, {\it Generalised Space-time and Gauge Transformations},
JHEP 1408 (2014) 050,  arXiv:1403.6395.  
\item{[19]} D. Steele and P. West, {\it E11 and Supersymmetry}, JHEP 1102 (2011) 101,  arXiv:1011.5820.
\item{[20]} S. de Buyl, M. Henneaux and  L. Paulot, {\it Extended E8
Invariance of 11-Dimensional Supergravity} JHEP {\bf 0602} (2006) 056
{\tt hep-th/05122992};  T. Damour, A. Kleinschmidt and  H. Nicolai {\sl
Hidden symmetries and the fermionic sector of eleven-dimensional
supergravity} Phys. Lett. B {\bf 634} (2006) 319 {\tt hep-th/0512163}; S. de Buyl, M. Henneaux and  L. Paulot {\sl Hidden
Symmetries and Dirac Fermions}Ê Class. Quant. Grav. {\bf 22} (2005) 3595
{\tt hep-th/0506009}. 
\item{[21]} N. Gromov, A Tumanov and P. West, to be published. 
\item{[22]} P. West, {\it The IIA, IIB and eleven dimensional theories and their common $E_{11}$ origin}, Nucl.Phys.{\bf B693} (2004 ) 76-102,  arXiv:hep-th/0402140.

\end